\documentclass{article}
\pdfoutput=1 
\usepackage{arxiv}
\usepackage{caption}
 \usepackage{multirow}
\usepackage{enumitem,kantlipsum}
\usepackage{amssymb}
\usepackage{mathtools}
\usepackage{multirow}
\usepackage{float}
\restylefloat{table}
\usepackage[utf8]{inputenc} % allow utf-8 input
\usepackage[T1]{fontenc}    % use 8-bit T1 fonts
\usepackage{hyperref}       % hyperlinks
\usepackage{url}            % simple URL typesetting
\usepackage{amsfonts}       % blackboard math symbols
\usepackage{nicefrac}       % compact symbols for 1/2, etc.
\usepackage{microtype}      % microtypography
\usepackage{lipsum}		% Can be removed after putting your text content
\usepackage{graphicx}
\usepackage{natbib}

\usepackage{doi}
\usepackage{amssymb}% http://ctan.org/pkg/amssymb
\usepackage{pifont}% http://ctan.org/pkg/pifont

\usepackage{graphicx}
\usepackage{lscape}
\title{Advances on the classification of radio image cubes}

\author{ \href{https://orcid.org/0000-0002-4197-1837}{\includegraphics[scale=0.06]{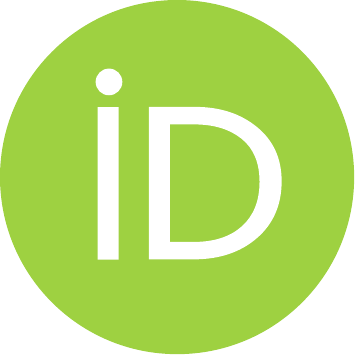}\hspace{1mm}Steven Ndung'u} \\
	Bernoulli Institute for Mathematics, Computer Science and Artificial Intelligence,\\
	University of Groningen,  \\  Groningen, The Netherlands\\
 \texttt{s.n.machetho@rug.nl} \\
	 \AND
	 Trienko Grobler \\
	 University of Stellenbosch, \\
	 Stellenbosch, South Africa, \\
	 \texttt{tlgrobler@sun.ac.za} \\
	 \And
	 Stefan J. Wijnholds\thanks{Part of this work is supported by the Foundation for Dutch Scientific
Research Institutes.} \\
	 ASTRON, \\
	 Dwingeloo, The Netherlands \\
	 \texttt{wijnholds@astron.nl} \\
	 \And
	 Dimka Karastoyanova \\
	 University of Groningen,  \\  Groningen, The Netherlands\\
	 \texttt{d.karastoyanova@rug.nl } \\
  \And
	 George Azzopardi \\
	 University of Groningen,  \\  Groningen, The Netherlands\\
	 \texttt{g.azzopardi@rug.nl} \\
}

\renewcommand{\shorttitle}{\textit{arXiv} Template}

\hypersetup{
pdftitle={Advances on the classification of radio image cubes}
}

\begin{document}
\maketitle

\begin{abstract}
Modern radio telescopes will daily generate data sets on the scale of exabytes for systems like the Square Kilometre Array (SKA). Massive data sets are a source of unknown and rare astrophysical phenomena that lead to discoveries. Nonetheless, this is only plausible with the exploitation of intensive machine intelligence to complement human-aided and traditional statistical techniques. Recently, there has been a surge in scientific publications focusing on the use of artificial intelligence in radio astronomy, addressing challenges such as source extraction, morphological classification, and anomaly detection. This study presents a succinct, but comprehensive review of the application of machine intelligence techniques on radio images with emphasis on the morphological classification of radio galaxies. It aims to present a detailed synthesis of the relevant papers summarizing the literature based on data complexity, data pre-processing, and methodological novelty in radio astronomy. The rapid advancement and application of computer intelligence in radio astronomy has resulted in a revolution and a new paradigm shift in the automation of daunting data processes. However, the optimal exploitation of artificial intelligence in radio astronomy, calls for continued collaborative efforts in the creation of annotated data sets. Additionally, in order to quickly locate radio galaxies with similar or dissimilar physical characteristics, it is necessary to index the identified radio sources. Nonetheless, this issue has not been adequately addressed in the literature, making it an open area for further study.
\end{abstract}

\section{Introduction}

Radio astronomy has seen an accelerated and exponential data eruption in the last two decades. Future radio telescopes like the Square Kilometre Array (SKA) will generate data sets on the scale of Exabytes. This will be one of the largest known big data projects in the world \citep{galaxies6040120}. The low-frequency instrument SKA-LOW will be located in Australia while the mid-frequency instrument SKA-MID will be located in South Africa. SKA-LOW will have a peak real-time data rate of 10 TB/s \citep{10.1117/1.JATIS.8.1.011024}, while SKA-MID will have a peak real-time data rate of 19 TB/s \citep{10.1117/1.JATIS.8.1.011021}. Other similar projects currently contributing to data-intensive research in astronomy that form the baseline/pathfinder to SKA include MeerKAT\footnote{https://www.sarao.ac.za/gallery/meerkat/}, which generates raw data at 2.2 TB/s \citep{booth2012overview}, the Murchison Widefield Array (MWA)\footnote{https://www.mwatelescope.org} with a data rate of  $\sim$300 GB/s \citep{lonsdale2009murchison} and the LOw-Frequency ARray (LOFAR) generating raw data at the rate of 13 TB/s \citep{refId0}. Astronomy has thus become a very data-intensive field with multi-wavelength and multi-messenger capabilities \citep{An2019ScienceOA}. These high data rates necessitate the automatic processing  of the data using computer intelligence. This motivates the need to assess the recent developments of computer intelligence applications within the field.

With the Evolutionary Map of the Universe (EMU) generating up to $\sim$70 million radio sources \citep{2011PASA...28..215N} and with the SKA expected to discover more than 500 million radio sources \citep{https://doi.org/10.48550/arxiv.1412.6076}, computer-aided applications are unavoidable. This has resulted in an increase in the number of scientific publications using machine/deep learning to detect and classify the radio sources. In the last five years, there has been successful proliferation of machine intelligence applications, owing to the availability of highly curated and annotated data catalogs Table~\ref{tab:datatable}).
Interestingly, publications on morphological classification have been on the incline, introducing novel and diverse machine/deep learning techniques to the radio astronomy field. This coupled with the above-mentioned progress and challenges have been our motivation to write this survey devoted to exploring the recent advancement in the classification of radio image cubes. Furthermore, other applications like anomaly/outlier detection, source extraction, and image retrieval will be discussed.

\begin{figure}[t]
\centering
\footnotesize
\includegraphics{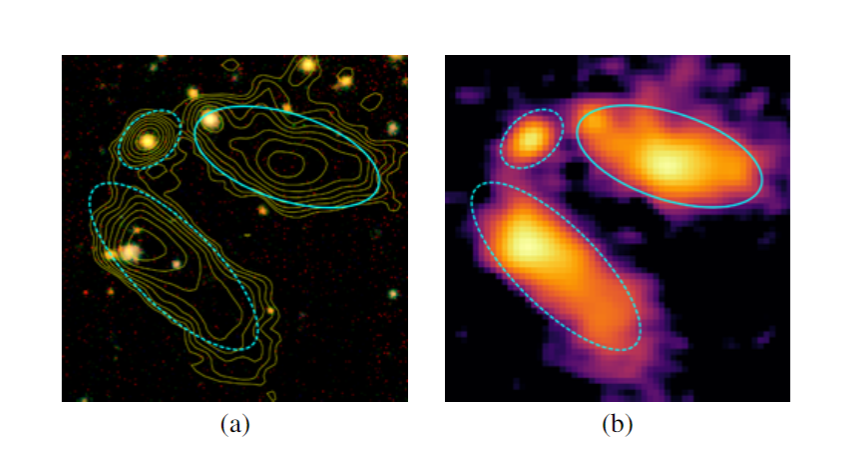}
 \caption{ An astronomical image as obtained from an optical and a radio telescope:  (a) the Legacy telescope (optical) $R$-band intensity, and (b) the LoTSS-DR2 stokes $I$ intensity. Source: Public LOFAR Galaxy Zoo: LOFAR. This is a typical example of a bent type galaxy.}
    \label{fig:image}
\end{figure}

Morphological classification is a crucial aspect of radio astronomy, as it allows scientists to understand the physical properties and characteristics of celestial objects based on their form and structure. Additionally, automated morphological analysis of large radio images can be a source of rare astrophysical phenomena, leading to serendipitous discoveries \citep{ray2016discovering}. This classification will focus on radio astronomy, which has played a very fundamental role in stimulating and spurring discoveries in the fields of cosmology, astrophysics, and telecommunications \citep{burke2019introduction}. Radio astronomy allows us to study celestial objects and phenomena at wavelengths that are not visible in the optical spectrum, providing unique insights into the universe. For instance, radio image cubes are supplemented by data obtained from other portions of the electromagnetic spectrum for cross-identification to help tackle fundamental scientific challenges. Fig.~\ref{fig:image}, obtained from the public LOFAR Galaxy Zoo: LOFAR project\footnote{https://www.zooniverse.org/projects/chrismrp/radio-galaxy-zoo-lofar}, illustrates this cross-identification process on an optical and a radio image of the same celestial object. These studies can help us better understand the physical processes at work in the universe and the diverse objects it contains \citep{burke2019introduction}. 

% \begin{figure}[t]
%     \centering
%     \footnotesize
%     \begin{tabular}{cc}
%         \includegraphics[width=4cm, height=4cm]{image1.png} &
%         \includegraphics[width=4cm, height=4cm]{image2.png} \\
%         (a) & (b) \\
%     \end{tabular}
%      % \includegraphics[width=8.5cm]{image_cubesv3.PNG} \\
%      % \begin{minipage}{4.25cm}\centering(a)\end{minipage}
%      % \begin{minipage}{4.25cm}\centering(b)\end{minipage}
%     \caption{ An astronomical image as obtained from an optical and a radio telescope:  (a) the Legacy telescope (optical) $R$-band intensity, and (b) the LoTSS-DR2 stokes $I$ intensity. Source: Public LOFAR Galaxy Zoo: LOFAR. This is a typical example of a bent type galaxy.}
%     \label{fig:image}
% \end{figure}

\subsection{Key challenges in radio astronomy}

In recent years, computer intelligence has been extensively applied to automate daunting manual and challenging tasks in radio astronomy. Some of the main areas that have experienced revolution and notable progress are telescope performance monitoring and the processing/transformation of visibility and image cube data. In modern telescopes, the demand for high-resolution observations and efficiency is very high, hence, the necessity of spontaneous real-time system health checks. To achieve this, machine learning algorithms are exploited \citep{10.1093/mnras/staa3087}. In \citet{10.1093/mnras/staa1412}, machine learning algorithms have demonstrated the capability to reliably detect, flag, and report system issues with above 95\% accuracy. This substantially mitigates the risk of failures while at the same time maintaining the peak performance of the telescopes. During the data curation stage in the visibility domain, machine learning techniques are used to automate the process of detection and correction of errors occurring in recorded data, while simultaneously removing outliers in the data sets \citep{Yatawatta_2021}. Furthermore, they are applied in the identification and extraction of radio frequency interference (RFI) - unwanted noise (signals) - which are produced by telecommunication technologies and other man-made equipment \citep{10.1093/mnras/stac570}. These kinds of signals and errors would degrade the quality of the data if not removed. 

In the image domain, the process of calibration relies heavily on the optimal fine-tuning of calibration parameters in the raw data processing pipelines.
Reinforcement learning is applied to automate the process of selecting and updating calibration parameters \citep{Yatawatta_2021}. This process is  a  tedious task due to the high number of calibration parameters that must  be tuned for telescopes with large fields of view \citep{5355494}. Moreover, astronomy has experienced a proliferation in the application of artificial intelligence in astronomical radio images to explore and address fundamental scientific challenges. The major areas of research in radio astronomy include: extraction and finding of radio sources such as point-like sources and extended sources \citep{https://doi.org/10.48550/arxiv.1910.03631,10.1007/978-3-030-89691-1_38}; classification  of the celestial objects based on their morphological features \citep{lukic2018radio,10.1093/mnras/sty2646}, spearheading  the advancement in the discovery of rare celestial objects such as pulsars, supernovas, quasars, and galaxies with unique and extraordinary  morphologies \citep{Mostert_2021}; and the retrieval of galaxies with similar  morphological characteristics \citep{AbdElAziz2017AutomaticDO}.

Generally, computer-aided systems have resulted in a paradigm shift in the capacity, capability, and rate at which immense and complex astronomical data is exploited relative to traditional methods. This has been further boosted by high computing, software, and hardware improvements - playing a critical role in the automation of the research processes in modern astronomy. Big data, however, still presents challenges due to its complexity, and the computational resources and execution times that are required by such data sets.

The rest of the paper is structured as follows: section \ref{sectn:radioastronomy}, provides a brief background on radio astronomy. Section \ref{sectn:survey_methodology} presents the approach followed to retrieve the relevant papers for this review. Section \ref{sectn:adoption_ci} provides a detailed review of the adoption of machine/deep learning algorithms in morphological classification. Section \ref{sectn:discussion} highlights the opportunities, challenges and future trends foreseen in the field of radio astronomy and finally, Section \ref{sectn:conclusion} presents a summary of the paper, highlighting the major insights from the review paper. \vfill

\section{Background}
\label{sectn:radioastronomy}
\subsection{Radio telescopes}

\begin{figure}[t]
\centering
\footnotesize
\includegraphics[width=\textwidth]{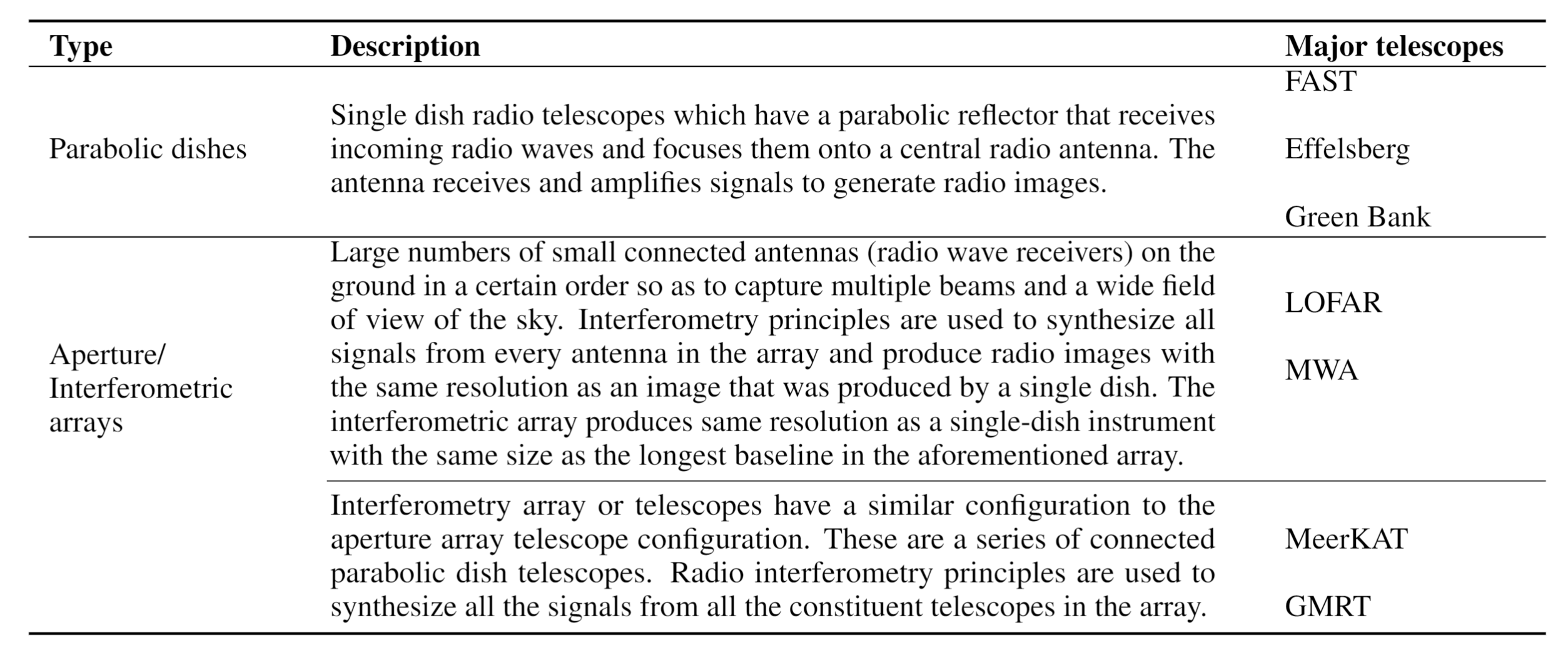}
\caption{Type and major radio telescopes of both the parabolic dishes and aperture arrays.}
\label{tab:radio_telescopes}
\end{figure}

\begin{figure}[t]
\centering
\footnotesize
\includegraphics{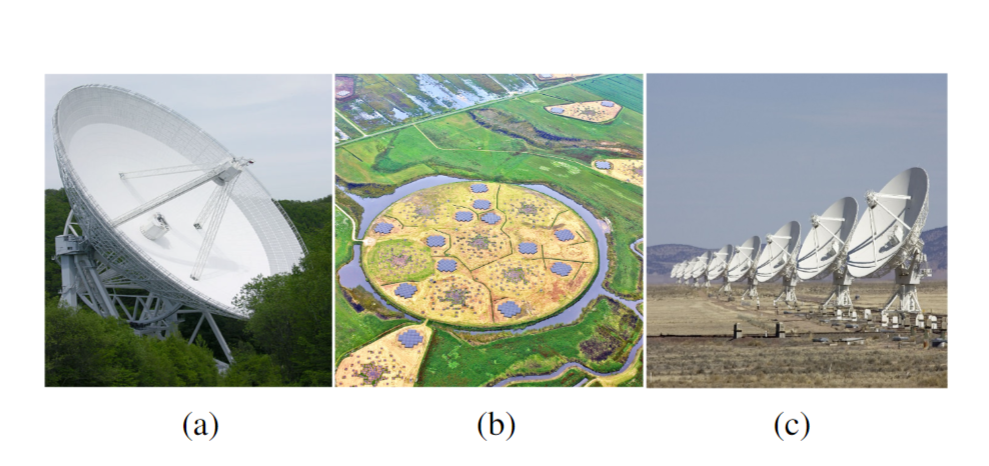}
   \caption[Parabolic dish and aperture array radio telescope]{Radio telescopes: a) Effelsberg radio telescope single parabolic dish, b) LOFAR antennas, and c) the Karl G. Jansky Very Large Array (VLA) telescope array.}
    \label{fig:telescopes}
\end{figure}

Radio telescopes are specialised astronomical instruments that detect and receive  very weak radio emissions radiated by extraterrestrial sources, for example, galaxies, planets, nebula, stars, and quasars. Radio telescopes can  either be single parabolic dishes, such as the Five hundred meter Aperture Spherical Telescope (FAST) in China or a number of inter-connected telescopes/antennas, namely the Giant Metrewave Radio Telescope (GMRT) in India and LOFAR in the Netherlands (Table~\ref{tab:radio_telescopes} and Fig.~\ref{fig:telescopes}).

Angular resolution and sensitivity are fundamental aspects to consider in a telescope. While angular resolution refers to the ability of a telescope to clearly differentiate radio sources observed in the sky, sensitivity is the measure of the weakest radio source emissions detected over the random background noise (the  flux density of celestial objects). Sensitivity is a product of several factors, namely signal coherence and processing efficiency, collecting aperture/dish area, along with receiver noise levels \citep{10.1117/1.JATIS.8.1.011021}. With high resolution and sensitivity, astronomers are able to clearly resolve between celestial objects and in doing so reveal more details of far faint stars and galaxies. The high angular resolution and sensitivity of radio telescopes have greatly boosted the acquisition of high resolution images through the next generation of wide-field radio surveys. For instance, LOFAR achieves a sensitivity of $\sim$100µJy/beam and a resolution of $\sim $6$''$ which enables it to detect sources that are faint and have small angular scales with a high resolution \citep{Shimwell_2022}.

% \begin{figure}[t]
%     \footnotesize
%     \centering
%     \includegraphics[width=8.5cm]{telescopesv3}
%     \\
%     \begin{minipage}{2.6cm}\centering(a)\end{minipage}
%     \begin{minipage}{2.7cm}\centering(b)\end{minipage}
%     \begin{minipage}{2.6cm}\centering(c)\end{minipage}
%     \caption[Parabolic dish and aperture array radio telescope]{Radio telescopes: a) Effelsberg radio telescope single parabolic dish, b) LOFAR antennas, and c) the Karl G. Jansky Very Large Array (VLA) telescope array.}
%     \label{fig:telescopes}
% \end{figure}

\subsection{Radio galaxies}

\begin{figure}[t]
\centering
\footnotesize
\includegraphics[width=\columnwidth]{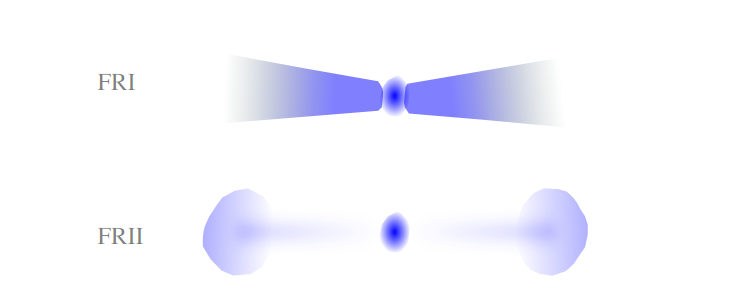}
\caption{A typical Fanaroff Riley I \& II classification of radio galaxies}
\label{fig:FRI_II}
\end{figure}

Radio galaxies are extensive astrophysical objects of radio emissions created by active supermassive black holes which form extended structures called jets and lobes. \cite{fanaroff1974morphology} proposed a seminal radio galaxy classification into two major families characterised by the distribution of luminosity of their extended radio emission. The first family is composed of centre-brightened (bright core) with one  or two lobes. They have brightened cores extending to the lobes; exuding a decaying luminosity from the core. They are called Fanaroff \& Riley I (FRI) galaxies. The second family is composed of edge-brightened lobes separated by a core at the center (the luminosity of the lobes decays as you move towards the center). They are referred to as Fanaroff \& Riley II (FRII) galaxies (Fig.~\ref{fig:FRI_II}). Further examination of the morphological characteristics of FRI and FRII galaxies resulted in the identification of the narrow-angled tail (NAT) and wide-angled tail (WAT) \citep{rudnick1976head} radio source populations with bent jets. In recent years, Fanaroff \& Riley 0 (FR0) galaxies, which are compact point-like sources, were added to the radio galaxy classification \citep{Baldi2015APS}. They are approximately five times compared to the total number of FRI and FRII sources and therefore constitute the largest population of radio galaxies \citep{baldi2018fr0cat}. Other rare and minority classes of sources include Ring-shape, X-shape, W-shape, S-shape or Z-shape, Double Double, Tri-axial, and other Hybrid morphologies \citep{proctor2011morphological}.

\begin{figure}[t]
\centering
\footnotesize
\includegraphics[width=\columnwidth]{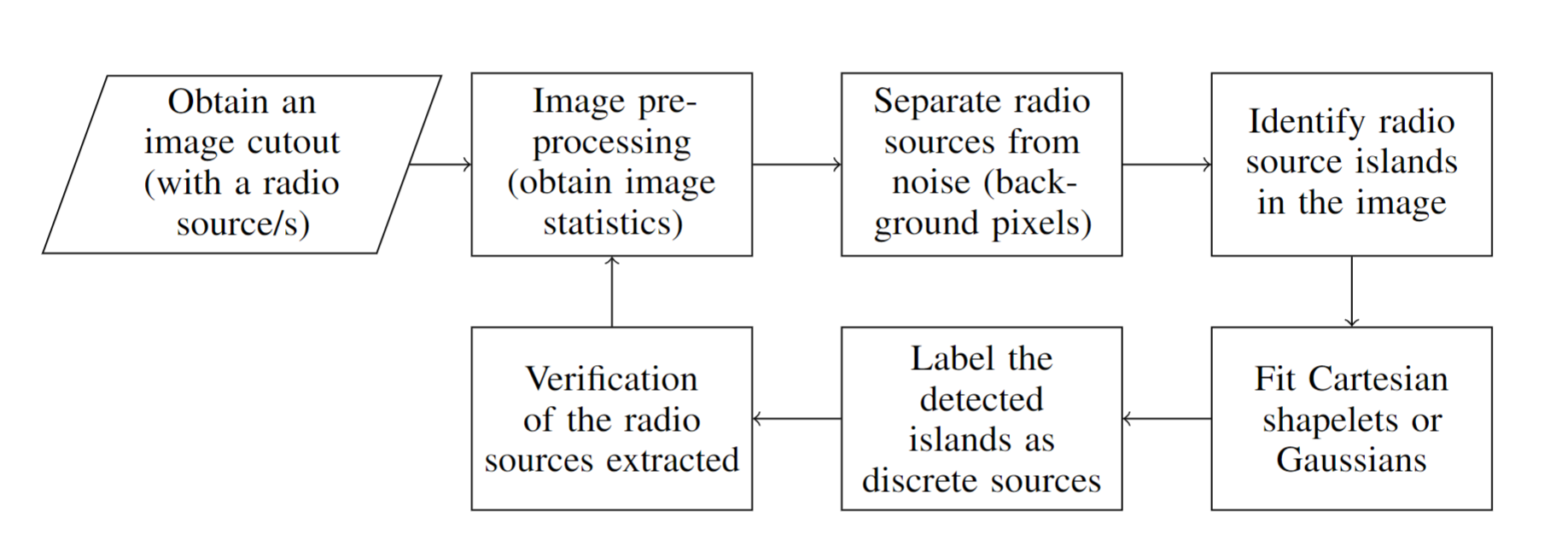}
\caption{The main steps illustrating the process of characterization and source extraction using PyBDSF.}
\label{fig:pybdsf_flowchart}
\end{figure}

% \begin{figure*}[t]
% \centering
% \footnotesize

% \tikzstyle{block} = [rectangle, draw, fill=white, text width=7em, text centered, minimum height=5em]
% \tikzstyle{io} = [trapezium, trapezium left angle=70, trapezium right angle=110,draw, fill=white, text width=6.7em, minimum height=4.4em, text centered]

% \begin{tikzpicture}[node distance = 3.2cm,auto] 

% \node [io] (dat_in) {Obtain an image cutout (with a radio source/s)};

% \node [block, right of=dat_in] (img_pr) {Image pre-processing (obtain image statistics)};

% \node [block, right of=img_pr] (sep) {Separate radio sources from noise (background pixels)};

% \node [block, right of=sep] (ident) {Identify radio source islands in the image};

% \node [block, below of=ident, yshift=1cm] (clus) {Fit Cartesian shapelets or Gaussians};

% \node [block, left of=clus] (val) {Label the detected islands as discrete sources};

% \node [block, left of=val] (res) {Verification of the radio sources extracted};

% \draw [->] (dat_in) -- (img_pr);

% \draw [->] (img_pr) -- (sep);

% \draw [->] (sep) -- (ident);

% \draw [->] (ident) -- (clus);

% \draw [->] (clus) -- (val);

% \draw [->] (val) -- (res);

% \draw [->] (res) -- (img_pr);

% \end{tikzpicture}

% \caption{The main steps illustrating the process of characterization and source extraction using PyBDSF.}
% \label{fig:pybdsf_flowchart}
% \end{figure*}

\subsection{Data management}
In data-centric fields such as astronomy, data management standards of the archived data are essential in conduit of knowledge discovery and innovation. They increase the rate of adoption of scientific discovery, knowledge integration and reuse in the wider community of researchers. The data management practices adopted must by design and implementation follow the FAIR (Findable, Accessible, Interoperable and Reusable) principles \citep{wilkinson2016a}. The system should allow easy data access, search, tagging, retrieval, and replication in an efficient and transparent way. This leads to seamless integration and  will allow global collaborations with other projects with similar data programs/systems.

Large radio astronomy facilities in the world store their data in either raw, calibrated/intermediate (for instance, VLA and LOFAR) or science-ready archives (for instance, ASKAP\footnote{https://www.atnf.csiro.au/projects/askap/index.html} and MeerKAT ) \citep{https://doi.org/10.48550/arxiv.2012.09273}. Some projects share their visibility data publicly via project-specific web interfaces\footnote{http://tdc-www.harvard.edu/astro.data.html}. Additionally, over the last few years, commendable progress in implementing FAIR principles in the field of astronomy has occurred due to the International Virtual Observatory Alliance (IVOA). It has been at the forefront of coordinating the integration of all the world's astronomy data into a federated system and has developed a standard set of protocols and specifications to be followed in astronomical data management \citep{https://doi.org/10.48550/arxiv.2012.09273}. IVOA enhances data interoperability across global astronomical data providers. Moreover, a case study conducted by the Australian All-Sky Virtual Observatory demonstrated that the implementation of the recommended IVOA standards and protocols results in \emph{almost} FAIR data \citep{https://doi.org/10.48550/arxiv.2203.10710}.

\subsubsection{Data annotation}
Finding, extraction, and characterization of radio sources which are typically galaxies containing an active galactic nucleus (AGN) or star-forming galaxy (SFG) and other celestial objects form the basis of the exploitation of radio surveys for scientific purposes. The data annotation mainly entails recovering the radio sources' delineation, position, estimated size, peak surface luminosity brightness, and providing labels and descriptions as per their morphological structure. The most reliable and accurate approach to annotating radio sources is a manual visual inspection of the images by radio astronomers. However, manual inspection by astronomers is limited due to the number of experienced astronomers dedicated to this task and also considering the size of the data.

\begin{figure}[t]
\centering
\footnotesize
\includegraphics{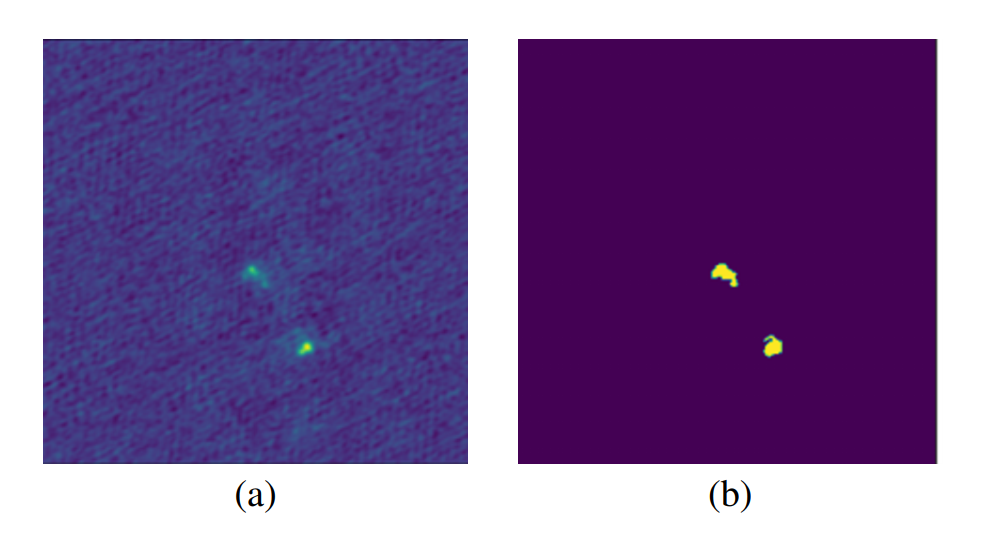}
\caption[Segmentation examples extracted by the pyBDSF software]{a) Original input image (with sources to be extracted) and b) two-component compact sources output as identified and extracted by the PyBDSF software.}
    \label{fig:pybdsf_extraction}
\end{figure}

Inspecting and characterizing radio sources is a difficult, costly, and time-consuming process. This has led to extensive development of statistical rule-based algorithms and methodologies for source extraction based on Cartesian shapelets, computer vision, Bayesian, and Gaussian methods. It has resulted in tools  such as the Python Blob Detector and Source-Finder (PyBDSF) \citep{2015ascl.soft02007M}, BLOBCAT, \citep{10.1111/j.1365-2966.2012.21373.x} and Aegean \citep{10.1111/j.1365-2966.2012.20768.x}. PyBDSF, for instance,  works based on the following algorithm, which is summarised in Fig.~\ref{fig:pybdsf_flowchart}: i) perform image pre-processing procedures and obtain image statistics, ii) determine a threshold value that separates the radio sources and the background noise pixels in the image, iii) with the background root mean square and mean values of the images, neighbouring islands of radio source emissions are identified, iv) the identified islands are fitted with multiple Cartesian shapelets or Gaussians to check if they are acceptable, and finally v) the Gaussians fitted within an identified/detected island are labeled and grouped into discrete sources. Additionally, Fig.~\ref{fig:pybdsf_extraction} shows an example of a two-component extended source extracted using PyBDSF. The study in \citet{Hopkins_2015} concludes that while these source finders are excellent for detecting compact sources, they suffer from insufficient robustness in the extraction of extended or diffuse sources. 

% \begin{figure}[t]
%     \footnotesize
%     \centering
%      \begin{tabular}{cc}
%         \includegraphics[width=3.5cm, height=3.5cm]{pybdsf_image1.png} &
%         \includegraphics[width=3.5cm, height=3.5cm]{pybdsf_image2.png} \\
%         (a) & (b) \\
%     \end{tabular}
%     % \includegraphics[width=8cm]{pybdsf} \\    
%     %  \begin{minipage}{3.9cm}\centering(a)\end{minipage} \begin{minipage}{3.75cm}\centering(b)\end{minipage} \\
%     \caption[Segmentation examples extracted by the pyBDSF software]{a) Original input image (with sources to be extracted) and b) two-component compact sources output as identified and extracted by the PyBDSF software.}
%     \label{fig:pybdsf_extraction}
% \end{figure}
\subsubsection{Data formats}
The most widely adopted community standard data formats in the field of astronomy include FITS (Flexible Image Transport System) \citep{Pence2010DefinitionOT}, Hierarchical Data Format (HDF5)\footnote{https://www.hdfgroup.org/}, Extensible N-Dimensional Data Format (NDF) \citep{2014ascl.soft11023W}, MeasurementSet (MS) \citep{VANDIEPEN2015174},  FITS-IDI \citep{greisen2011fits}, and UVFITS \citep{greisen2012aips}. The various formats have different strengths and weaknesses when it comes to the different data processing tasks, namely recording, transferring and archiving. For example, HDF5 format is excellent for  data processing, transfer, and storage relative to other formats as it supports parallel I/O, distributed and data chunking mechanisms, and data compression which is very important in the era of big data \citep{https://doi.org/10.48550/arxiv.1411.0507}.

%\clearpage
\begin{figure}[t]
\centering
\footnotesize
\includegraphics[width=\textwidth]{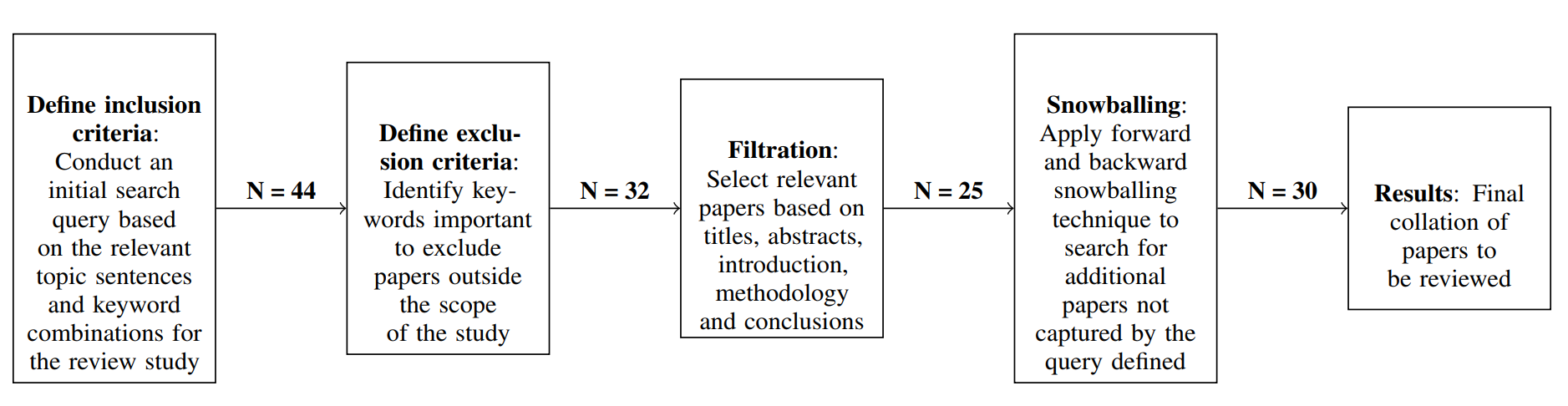}
\caption{ The protocol followed to identify relevant articles for this survey. N represents the number of papers selected after each selection stage.}
\label{fig:surveymethodology}
\end{figure}

\subsubsection{Commonly used catalogs}
\begin{figure}[t]
\centering
\footnotesize
\includegraphics[width=\textwidth]{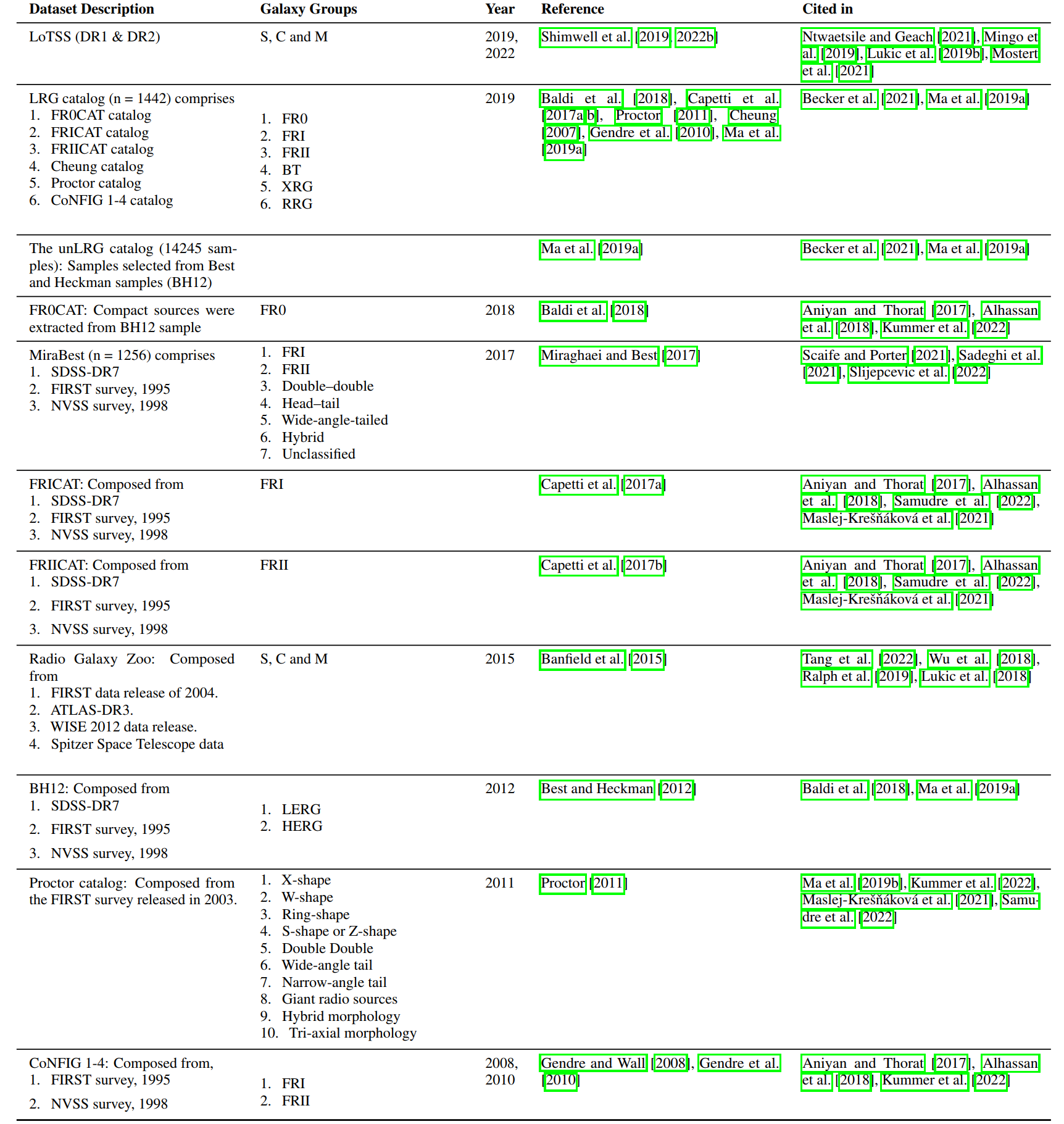}
\caption{Commonly used data sets for morphological and anomaly detection. Abbreviations are defined in the Appendix.}
\label{tab:datatable}
\end{figure}

The compilation of annotated data catalogs that are publicly available and accessible is an important contribution to the promotion of the development of research in morphological classification of radio galaxies. Catalogs were compiled with different objectives such as detailed exploration, comparison and examination of a given population of galaxies \citep{baldi2018fr0cat, 10.1093/mnras/stx007}, provision of large and comprehensive labelled data sets for mining radio galaxy morphologies \citep{10.1111/j.1365-2966.2010.16413.x, proctor2011morphological} and creating a representative and balanced catalogs encompassing different classes of radio galaxies \citep{aniyan2017classifying, ma2019machine}. Owing to the varied aims and different procedures of sample selection in developing the catalogs, the number of radio morphological classes per data set is different. For example, some catalogs contain a single class \citep{baldi2018fr0cat, capetti2017fricat, capetti2017friicat},  two classes \citep{10.1111/j.1365-2966.2012.20414.x, gendre2008combined, 10.1111/j.1365-2966.2010.16413.x}, or more  \citep{10.1093/mnras/stx007, ma2019machine, proctor2011morphological}. Additionally, the catalogs are derived from various radio telescope surveys with different levels of luminosity. Table~\ref{tab:datatable} 
summarises the commonly used data sets in machine/deep learning applications of radio astronomy.

\section{Survey methodology}
\label{sectn:survey_methodology}
The motivation of this survey paper is to give an account of the recent progress of
computer intelligence in morphological classification in radio image data, with a focus on the last five years that have seen substantial progress in deep learning paradigms. Besides the core topic mentioned above, supplementary challenges like image annotation, data management, anomaly detection, and scalability are also considered to some extent.
%\vfill
%\clearpage
Web of Science\footnote{https://www.webofscience.com/} and NASA's Astrophysics Data System\footnote{https://ui.adsabs.harvard.edu/} databases were used to retrieve relevant literature papers for the study and the results cross-checked on Google Scholar\footnote{https://scholar.google.com/} database. These databases offer advanced search capabilities and comprehensive coverage of high-quality journal articles across various disciplines, particularly in the areas of Computer Science and Astronomy, which are the focus of our research.

\begin{figure}[t]
\centering
\footnotesize
\includegraphics[width=0.9\textwidth]{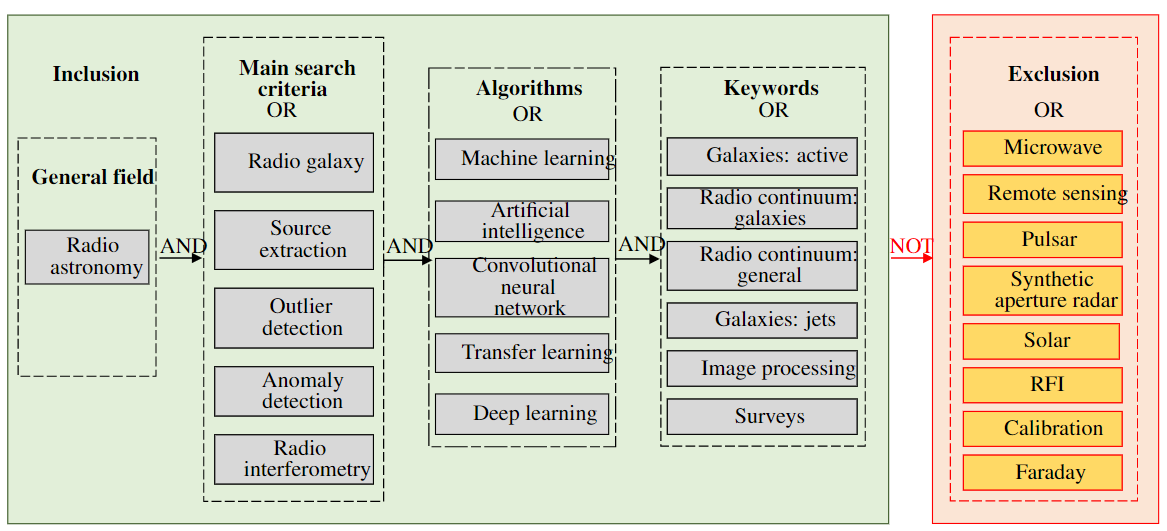}
\caption{A schematic study design process of exclusion and inclusion criteria adopted for the retrieval of the relevant articles considered in this survey.}
\label{fig:schematic_flowchart_LRP_v2}
\end{figure}

We aimed to achieve fair and representative sample papers from the large pool of published papers over the last five years. The search strategy protocol adopted is outlined in Fig.~\ref{fig:surveymethodology}, \citep{wee2016write}. Furthermore, Fig.~\ref{fig:schematic_flowchart_LRP_v2} illustrates the schematic study design of inclusion and exclusion criteria that were used. A total of 44 papers were retrieved from the initial query. Thereafter, an exclusion criterion was introduced to filter out papers in the fields of remote sensing and those in the field of radio astronomy but covering RFI, pulsars, solar and microwaves, as we consider them beyond the scope of our review. After retrieving relevant papers using refined queries on Table~\ref{tab:search_query}, we then applied the forward and backward snowballing technique of the obtained papers \citep{Wohlin2014GuidelinesFS}. This left us with a total of 30 papers. Notably, from the final selection of papers extracted, there was no review paper covering the scope of radio astronomy. The few available papers identified were in the wider field of astronomy, assessing the adoption and maturity of machine learning and deep learning in the field \citep{fluke2020surveying,wang2018computational}.

\begin{table}[t]
\caption{Search query used in Web of Science for the retrieval of relevant review papers. TS = Topic sentence and KS =  Keywords Plus. Quotation marks are used for exact matching.}
\label{tab:search_query}
\centering
\footnotesize
\begin{tabular}{p{\textwidth}}
\hline
\textbf{Web of Science Query} \\
\hline
Query = ((TS=("radio astronomy" OR "radio galaxy" OR "radio interferometry"  ) AND  TS=("radio" OR "anomaly" OR "outlier" OR "source extraction") AND TS=("*machine learning*" OR  "*convolutional neural network* "OR "*deep learning*" OR "*transfer learning*" OR "artificial intelligence*") OR KP=("galaxies:active", "radio continuum:galaxies", "radio continuum:general", "galaxies:jets","image processing", "surveys","galaxies:active", "radio continuum:galaxies", "radio continuum:general", "galaxies:jets","image processing", "surveys"')) NOT TS=(  "solar" OR "rfi" OR "pulsar" OR "remote sensing" OR "synthetic aperture radar" OR "microwave")) \\
\hline
\end{tabular}
\end{table}

 \begin{figure}[t]
\footnotesize
\centering
\includegraphics[scale=1.1]{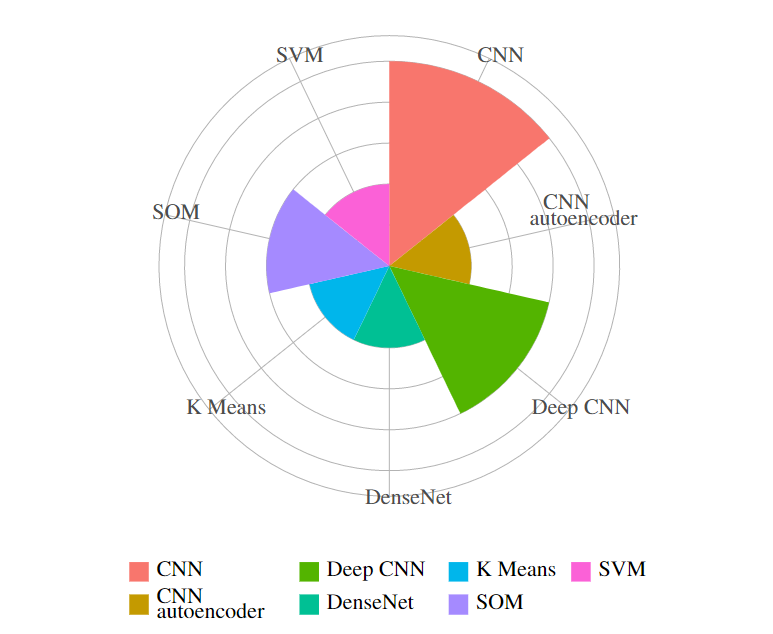}
 \caption{A Coxcomb chart illustrating the top seven most commonly used machine learning methodologies in radio astronomy in recent years. The quantity of papers belonging to each of the seven categories is equal to the number of concentric circles that overlap the respective segment.}
\label{fig:Coxcombchart}
\end{figure}

\begin{figure}[t]
\centering
\footnotesize
\includegraphics[width=\textwidth]{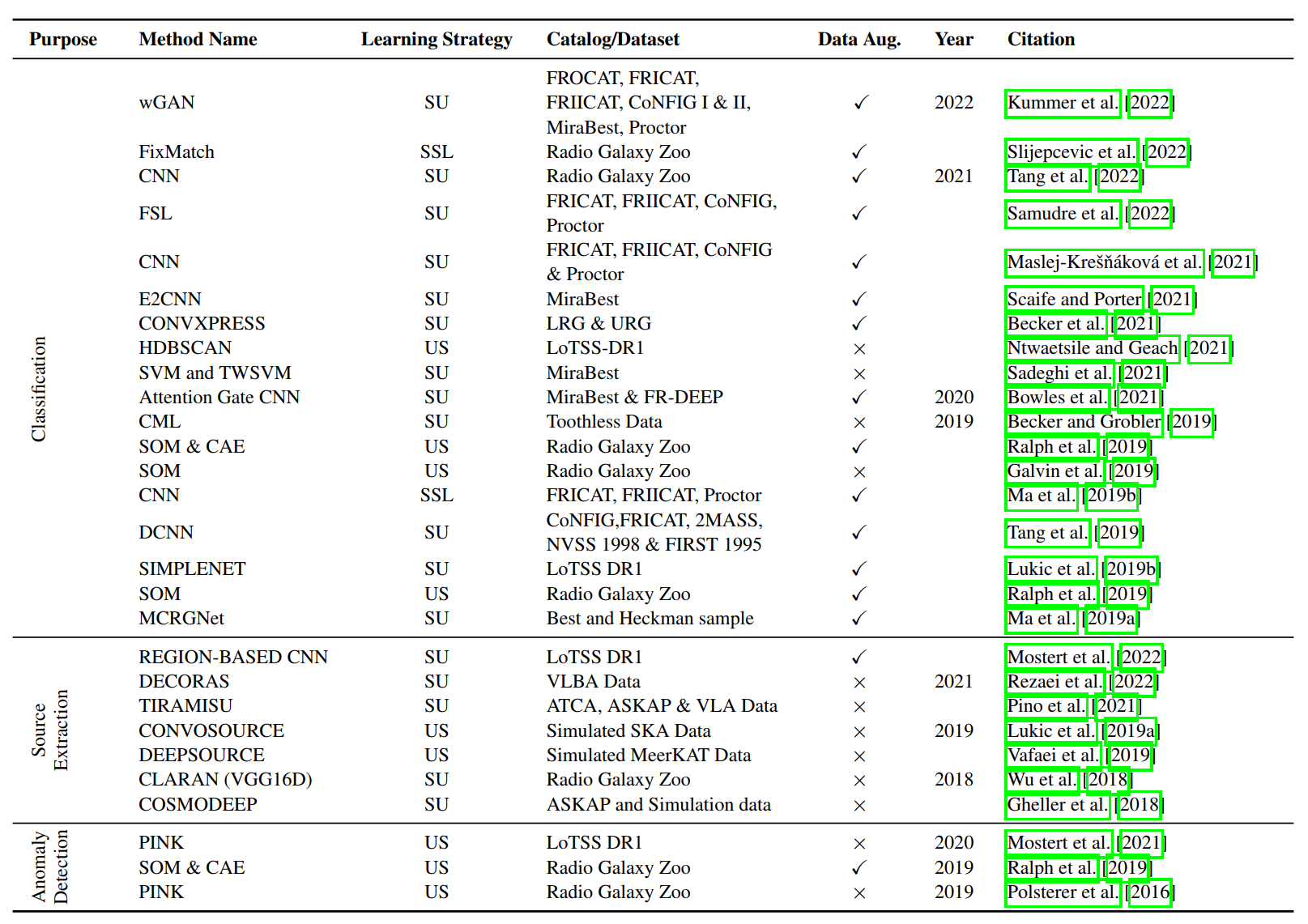}
\caption{Summary of classification, source extraction and anomaly detection papers. Abbreviations are defined in the Appendix.}
\label{tab:survey_papers}
\end{figure}

Table~\ref{tab:survey_papers} presents a high-level summary of the surveyed papers. The papers provide a wide range of machine/deep learning-based methods applied in the field of radio astronomy. In the coxcomb chart (similar to a pie chart) shown in Fig.~\ref{fig:Coxcombchart}, the radius of each circle segment is proportional to the number of papers it represents. Therefore, the radius is determined by the frequency of the methodology in the papers surveyed. It can be observed that the majority of the methodologies used are based on shallow and deep convolutional neural networks (CNNs). Radio astronomy has indeed adopted and adapted the latest innovative and novel methodologies such as deep CNNs and Transformers from the larger science community. This has consequently led to the development of massive data-driven intelligent pipelines, which have automated the rather inefficient historically manual process.

\newcommand{\cmark}{\ding{51}}%

\section{Adoption of computer intelligence in radio astronomy}

\label{sectn:adoption_ci}
The adoption of artificial intelligence in radio astronomy has led to a plethora of machine and deep learning applications in classification and segmentation tasks. This has been majorly attributed to the resurgence of artificial intelligence, resulting in the development of innovative and novel deep learning architectures such as CNNs (also known as ConvNets) due to the exploitation of high-resolution images. ConvNets are to some extent inspired by the biological functionality of the human visual cortex. They have become the de facto choice for many computer vision tasks.

\begin{figure*}[t]
 \centering
 \footnotesize
\includegraphics[scale=1]{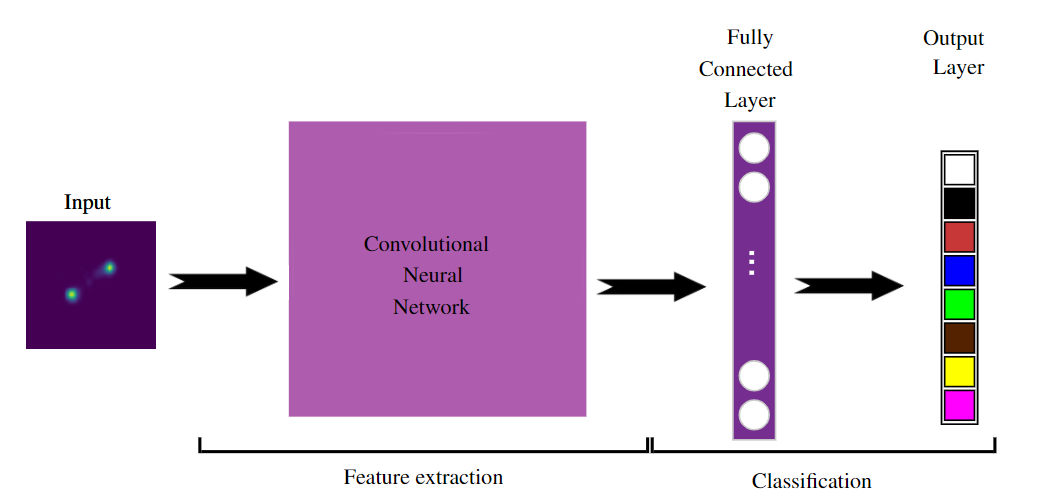}
\caption{The fundamental building blocks of a standard ConvNet.}
\label{fig:simple_cnn_structure}
\end{figure*}

A simple ConvNet is generally composed of a set of convolutional (multiple building blocks), and subsampling (pooling) layers followed by a fully connected layer as shown in Fig.~\ref{fig:simple_cnn_structure}. In addition, various linear and non-linear mapping functions and regulatory units are embedded in the structure (e.g activation functions, batch normalization, and dropout) to optimize its performance. CNN models are designed  to automatically and adaptively learn spatial features during training. The convolution and subsampling layers are focused on feature extraction while the fully connected layer maps the extracted features onto outputs. In the early layers of a CNN, simple features like edges are identified. Then, as the data progresses through the layers, more sophisticated features are determined. Notably, ConvNets classify images based on learned weights in the form of convolutional kernels obtained through the training process.

\begin{figure*}[t]
 \centering
  \footnotesize
\includegraphics[trim={0 3.1cm 5cm 5.5cm},clip,scale=0.33]{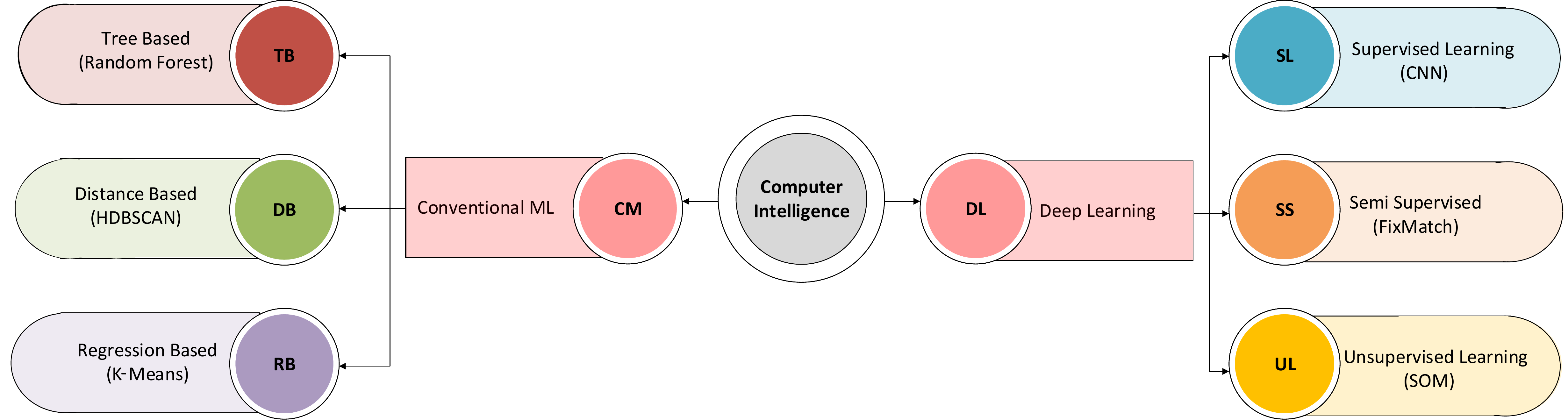}
 \caption{Computer intelligence methodologies applied in the classification of radio galaxies.}
\label{fig:CI_in_aadio_astronomy}
\end{figure*}

In the next section, we delve into a synthesis of the papers listed in Table~\ref{tab:survey_papers}.

\subsection{Morphological classification}
\label{lbl:analyses}

The generation of science-ready survey catalogs requires the classification of processed calibrated radio images  into various physical source categories such as galactic, extragalactic, AGN, and SF galaxies. The process of identifying and annotating such phenomena is very crucial in the preparation and release of science-ready products to the public for further scientific exploitation. Additionally, the process helps scientists to have a better comprehension of the Universe through exploring the fundamental laws of physics. Therefore, automating the process of visualization and the labeling of sources based on their morphological features is, therefore, critical in astronomy.

Broadly, morphological classification in radio astronomy entails grouping populations of  Fanaroff-Riley (FR) radio galaxies into compact (point-like) and extended sources (FRI, FRII, WAT, NAT, XRG - X-shaped radio galaxies, RRG - ringlike radio galaxies, along with others); the extended sources contain complex morphological structures with two or more components in a galaxy. The developed FR classification approaches utilize  either unsupervised, semi-supervised or supervised machine learning. Fig.~\ref{fig:CI_in_aadio_astronomy} illustrates the general taxonomical categorization of classification methods reviewed.

Using supervised learning, \cite{aniyan2017classifying} developed the first ConvNet model based on Alexnet CNN architecture (Toothless\footnote{https://github.com/ratt-ru/toothless}). Their model was evaluated on the Toothless\footnote{Toothless is a three-class radio galaxy data set composed of selected well-resolved FRI (178 samples), FRII (284 samples), and Bent-tailed (254 samples) sources.} data set achieving  accuracies of 95\%, 91\% and 75\% for Bent-tailed, FRI and FRII, respectively. Their work provided a baseline that clearly demonstrates the  potential of deep learning in classifying radio galaxies. Besides, the VGG-16 architecture \citep{7486599}$^*$\footnote{The symbol $^*$ is used on citations that are not part of the papers under review} was used in a semi-supervised way to classify radio galaxies and as such it leverages the large unlabelled data sets that are available \citep{ma2019classification}.

Unsupervised learning using methodologies like self-organizing maps were used by \citet{Polsterer_2016}, to construct radio morphologies based on similar/dissimilar characteristics of the Radio Galaxy Zoo project data \citep{banfield2015radio}. 
The authors proposed the Parallelized rotation and flipping INvariant Kohonen maps (PINK) approach, which does not require training data labels, and hence avoids any potential bias by inexperienced practitioners in the Radio Galaxy Zoo project \citep{banfield2015radio}. It only required human inspection and profiling of the resulting prototypes into known FR galaxy sources accordingly. 

While deep learning methodologies are seen to be dominant in the classification task as seen in Table~\ref{tab:survey_papers}, conventional machine learning techniques have also been explored in the classification of FR galaxies. \citet{9015881} compared the following methodologies: Nearest Neighbors \citep{peterson2009k}$^*$, Support Vector Machine (SVM) \citep{cortes1995support}$^*$, Radial Basis Function SVM \citep{DING202110121}$^*$, Gaussian Process Regression \citep{banerjee2013efficient}$^*$, AdaBoosted Decision Tree \citep{freund1997decision}$^*$, Random Forest \citep{breiman2001random}$^*$, Naive Bayes \citep{rish2001empirical}$^*$, Multi-layered Perceptron \citep{piramuthu1994classification}$^*$ and Quadratic Discriminant Analysis \citep{bose2015generalized}$^*$ in the classification of Fanaroff-Riley Radio Galaxies. \cite{9015881} used the Toothless data set excluding the bent-tailed radio sources in their implementation. A comparative analysis was performed between different conventional machine-learning algorithms on radio images. The Random Forest classifier was found to have the highest performance with an accuracy of 94.66\% \citep{9015881}. The study demonstrated that the derived morphological features from radio images are distinct and unique to radio galaxy classes. 

In order to comprehensively discuss the papers under review, we consider data processing pipelines and model architectures used in the research papers. Specifically, the methodological applications covered in this review are categorized into three major groups: model-centric approaches, data-centric approaches,  and weakly supervised approaches. This is motivated by the need to develop robust algorithms when limited annotated data is available or when massive amounts of unlabelled data can be utilized.

\subsection{Model-centric approach}

Research in computer intelligence predominantly dedicates resources and time to improving and optimizing machine learning algorithms. Developing novel model architectures has been witnessed in the space of deep learning. This has gradually been translated into the field of radio astronomy given it is a data-driven field.   

\subsubsection{CNN architectures}

Model architectures have been shown to  play a significant role in improving and increasing the generalization  of deep learning algorithms in classification problems. Therefore, we have seen progressive breakthroughs and applications of more complex architectures such as AlexNet \citep{krizhevsky2017imagenet}$^*$\citep{aniyan2017classifying}, VGG-16  \citep{ma2019classification,10.1093/mnras/sty2646}, and DenseNet \citep{8099726}$^*$ \citep{samudre2022data} in radio astronomy. The depth of the CNN architecture models are varied across different applications, depending on the required complexity. For instance, \citet{Lukic_2019} constructed four-layer (CONVNET4) and eight-layer (CONVNET8) convolutional networks, \citet{becker2021cnn} constructed eleven layers, \citet{aniyan2017classifying} constructed twelve layers, and \citet{2019MNRAS.488.3358T} constructed thirteen layers for classification of radio galaxies. According to a comparative analysis done on a capsule network, CONVNET4 and CONVNET8 on the LoTSS DR1 data set, it was observed that CONVNET8 outperformed CONVNET4 and a capsule network, though with a marginal difference \citep{Lukic_2019}.  The eight- and four-layer CNNs and the capsule network attained average precision scores of 94.3\%, 93.3\% and 89.7\%, respectively. The secret behind the increase in depth of the convolutional layers is that it augments the number of nonlinear functions and introduces additional feature hierarchies that optimize the classification function. Consequently, the deep networks tend to achieve higher performance compared to more shallow networks \citep{2019MNRAS.488.3358T}.

\subsubsection{Regularization techniques}

Overfitting has been one of the central challenges affecting the robustness of radio galaxy classification models. The availability of small labeled astronomical data sets for building the models remains to be a major contributor to the challenge. To address this, researchers have adopted regularization techniques during model building. This is aimed at allowing the models to maximally learn from the limited training data and achieve better generalization. One technique used is the random dropping out of weakly connected units (neurons) of CNN connections during training \citep{2019MNRAS.488.3358T,tang2022radio}. This approach is commonly referred to as dropout. Dropout helps to reduce parameter saturation during the training process preventing excessive co-adapting of the units. Moreover, to reduce covariance shift in the input data, the batch normalization technique is applied during model training \citep{2019MNRAS.488.3358T,tang2022radio}. This involves standardizing the feature maps such that the values are transformed to follow a Gaussian distribution (regularize the network). These regularization approaches reduce the chances that the network will succumb to the vanishing gradient problem and reduce the time that the network requires to converge.

\subsubsection{Specialized convolutional blocks}

The key thrust in the performance of ConvNets compared to other models is the continued construction and integration of innovative processing units and the embedding of newly designed novel convolutional blocks. In radio astronomy, there are several novel research efforts in this direction. 

Attention gates are convolutional blocks that are analogous to the visual system of humans to efficiently prioritize localized salient features in an object in order to contextualize and identify it. \citet{bowles2021attention} implemented novel convolutional filters that localize salient features while suppressing irrelevant information on the provided images, thus, resulting in predictions obtained directly from pertinent and contextualized feature maps. The attention-gate layers are integrated in the CNN architectural backbone. This approach was found to reduce the CNN model training parameters by 50\%  and improves the interpretability of CNN models. It promotes explainable deep learning by using  attention maps that can be investigated to trace the root cause of misclassification in a model.  Despite the notable reduction in training parameters, the performance of the CNN architecture developed was equivalent to the state-of-the-art CNN applications in the literature. 

Group equivariant Convolutional Neural Networks (G-CNNs)  are convolution kernel filters that are embedded in the conventional CNN \citep{cohen2016group}. G-CNNs are aimed at supporting equivariance translation for a wider set of isometries (for example rotation and reflections) on the training data. By design, CNNs are constructed to be translation-equivariant of their feature maps, but this does not apply to other isometries such as rotation. This implies that G-CNNs allow preservation of group equivariance on augmented data - a common data-centric approach in deep learning model building. Thus, the increased data samples via rotational augmentations result in the same kernel (weight sharing) as they pass through the convolutional layers. This approach has been demonstrated to improve CNN architecture performance in the galaxy classification task using the MiraBest data set \citep{scaife2021fanaroff}.   

Other innovative ideas introduced to the standard convolutional architectures in radio astronomy include, multidomain multibranch CNNs, which allow the models to take multiple data inputs as opposed to single source images \citep{10.1093/mnras/sty1308,tang2022radio}.

\subsection{Data-centric approaches}

The quality and robustness of machine and deep learning algorithms are highly dependent on the quality of data. Quality entails the consistency, accuracy, completeness, relevance, and timeliness of the data. Principally, in order to improve the performance of the algorithms, data-centered approaches are paramount. The data (radio images) must be free from RFI noise and artifacts before calibration and processing. The data should not be ambiguous and each sample should belong to a definite radio galaxy class. Ideally, data must be highly curated.

In addition, to circumvent overfitting and simultaneously achieve high generalization accuracies, adequate data diversity on the training data set is a prerequisite. This aids in avoiding poor model performance when tested with real-world out-of-distribution data or covariate-shifted data.

\subsubsection{Data augmentation}
Data augmentation aims to increase the size and diversity of the training set. It is applied on the assumption that additional important information can be extracted from the insufficient data set available via augmentations. It has been widely espoused in radio galaxy classification to mitigate overfitting \citep{aniyan2017classifying,alhassan2018first,lukic2018radio}, to improve the performance of machine and deep learning models \citep{maslej2021morphological,kummer2022radio, lukic2018radio}, to address rotational invariance \citep{becker2021cnn}, to increase the size and the diversity of the training data \citep{aniyan2017classifying,alhassan2018first,becker2021cnn,ma2019machine},  and to address the class imbalance, especially for the minority classes among the radio galaxy population groups in the training data \citep{lukic2018radio}. There are different kinds of augmentation strategies. Two of these strategies are positional augmentation and color augmentation. Examples of the former include scaling, flipping, rotation, and affine transformation. Examples of the latter include brightness, contrast, and saturation \citep{10.1111/j.1365-2966.2012.20414.x,becker2021cnn,scaife2021fanaroff,slijepcevic2022radio}. Other augmentation approaches include up-sampling or oversampling of the minority class and generative adversarial networks \citep{kummer2022radio}. The literature attests to the fact that data augmentation is a data-centered strategy that can significantly improve model performance and result in models with improved generalization ability\citep{maslej2021morphological}.

\cite{maslej2021morphological} found that   improvement of model performance and capacity to generalise on out-of-distribution data was highly dependent on augmentation strategy that was employed. They found that brightness increase, vertical or horizontal flips, and rotations led to better performance while zoom, shifts, and decrease in the brightness of the images degraded model performance. Therefore, the process of finding an optimal data augmentation strategy in a project is non-trivial. A downside of data augmentation is that any inherent bias or data errors will be inherited by the augmented data. Nevertheless, this does not rule out the fact that  data augmentation is an important data-centric approach for both increasing minority data classes and improving model performance in the computer vision paradigm.

\subsubsection{Feature engineering}
Feature engineering is aimed at improving model accuracy in machine learning. It involves the process of careful selection based on domain knowledge, feature extraction, creation, manipulation, and transformation of the training data. The engineered features are targeted at providing the `precise physical properties' of the image data for model development. In radio galaxy classification, morphological features engineered include peak brightness, lobe size, number of lobes, and right ascension and declination \citep{9015881}. Moreover, feature descriptors that represent the texture of radio images via Haralick features\footnote{Haralick features are a set of thirteen non-parametric measures which are derived from the radio images based on the Grey Level Co-occurrence Matrix.} \citep{10.1093/mnras/stab271} and use Radial Zernike polynomials to extract image moments such as translation, rotation, that are scale-invariant \citep{2021AJ....161...94S}. 

Machine learning algorithms are applied on the features  engineered (compact representations of the radio images) for classification of radio galaxies. In this case, either supervised or unsupervised approaches are used, for example, Hierarchical Density Based Spatial Clustering of Applications with Noise (HDBSCAN) \citep{10.1093/mnras/stab271}, Random Forest (RF) \citep{9015881} and SVM \citep{2021AJ....161...94S}. Feature engineering has been shown to provide machine learning algorithms with features of high importance resulting in high performances, with accuracies above 95\%  \citep{2021AJ....161...94S}. However, the main drawback is that it requires domain expertise to design feature descriptors. Therefore, they may not be able to capture all the relevant information in the data.

\subsection{Weak supervision approaches}

In radio astronomy, most publicly available catalogs contain 10$^3$ radio galaxies. Moreover, the cost of labeling sufficiently large (in deep learning terms) radio astronomical data sets is very high. On the contrary, unlabelled catalogs consist of Petabytes of data (from a single survey). Hence, the essence of exploring algorithms and strategies with the capacity of leveraging the massive unlabelled public catalogs and/or exploiting the small annotated data sets available are paramount. 

The three weakly supervised methods, namely transfer learning, semi-supervised learning, and N-shot learning are discussed.  

\subsubsection{Transfer learning}
Transfer learning is a paradigm that reuses knowledge gained from pre-trained models on massive data sets to fine-tune them on other tasks, making it effective for scarce training data. In the context of classification of radio galaxies, transfer learning has been investigated and has contributed to improved accuracies compared to other methods, such as few-shot learning \citep{samudre2022data}. The pre-trained model's weights and biases provide the generic feature representations essential to the model for identifying low-level features (i.e, shapes and edges) of the objects. Then, the complementary complex features specific to the classification task at hand are learned by fine-tuning the last layers of the model using the available small labeled data set. The study by \citet{2019MNRAS.488.3358T} investigated whether it was possible to develop robust cross-survey identification machine learning algorithms that made use of the transfer learning paradigm. In their research, they used FIRST and NVSS survey data, which are characterized by high- and low-resolution images, respectively. They found that models pre-trained on high-resolution surveys (FIRST) can be effectively transferred with high accuracies of about  94\% (a case of 2 classes: FRI and FRII), to lower-resolution surveys (NVSS). However, the converse was observed not to be true. 

Similarly, transfer learning on radio galaxy classification has been shown to achieve high performance even after extending the number of classes to more than two: FRI and FRII. \citet{Lukic_2019} used Inception ResNet model v2 \citep{szegedy2017inception} to classify three classes  (FRI, FRII, and Unresolved) from the LoTSS-DR1 data. Inception ResNet model v2 achieved an average accuracy of 96.8\%; the best performance compared to ConvNet-4, ConvNet-8 and Capsule Networks model architectures that they experimented with on the same data set. Additionally, a transfer learning method based on the Dense-net architecture  \citep{8099726}$^{*}$ was tested by \citet{samudre2022data}. They obtained a precision of 91.9\%, a recall of 91.8\% and an F$_{1}$ score of 91.8\% for the classification of compact, FRI, FRII, and Bent radio galaxies  with less than 3000 test samples \citep{samudre2022data}. Notably, transfer learning was observed to  converge faster compared to conventional CNN architectures. For instance, the model converged faster (10 fewer epochs on average) than other models such as ConvNet-4 \citep{Lukic_2019}.

\subsubsection{Semi-supervised learning}
Semi-supervised learning (SSL) lies between unsupervised and supervised learning, utilizing both annotated data samples and a large amount of unannotated data during training. Employing semi-supervised techniques for the radio galaxy morphological classification task has recently been gaining traction within the literature. The reason for this can be ascribed to the fact that there are large publicly available unannotated data sets that are available for use within the field of radio astronomy.

Concerted efforts have been dedicated to investigating the possibility to exploit these algorithms and conduct a comparative analysis of the performance with supervised machine learning \citep{ma2019classification,ma2019machine,slijepcevic2022radio}. \citet{ma2019classification} trained a semi-supervised model where they constructed a radio galaxy morphology classifier (autoencoder) from  the VGG-16 architecture. The autoencoder was pre-trained on a large unannotated data set of 18,000 radio galaxies from the BH12 catalog \citep{10.1111/j.1365-2966.2012.20414.x}. The pre-training of the modified VGG-16 architecture was aimed at updating its weight and bias parameters - allowing the model to learn the low-level morphological features of the radio galaxies (such as shapes and outlines). The pre-trained model was then fine-tuned with a small annotated data set of about 600 radio galaxies only. It was observed that the SSL strategy achieved high average precision and recall of 91\% and 90\%, respectively. % with supervised \citep{aniyan2017classifying} and transfer learning.
Similarly, the MCRGNet classifier (SSL model) was pre-trained on the unLRG (unlabelled radio galaxy) (14,245 samples) and fine-tuned on the LRG (labeled radio galaxy) (1442 samples) data sets \citep{ma2019machine}. The MCRGNet's average classification precision was 93\%. This was a better precision compared to the competing methods at the time. 

Another methodological approach used in SSL for radio galaxy classification is presented by \citet{slijepcevic2022radio}, which used the FixMatch algorithm \citep{sohn2020fixmatch}$^*$. In FixMatch's strategy, a weakly augmented (for instance, shift or flip data augmentation methods) unannotated image is first fed into a model and then used to generate a pseudo-label. Then, in a concurrent fashion, the same unannotated image under strong augmentations (for instance, brightness, translation, or contrast) is fed into a model to generate a prediction. Thirdly, using cross-entropy or a distance measure, such as Fréchet inception distance, the model is trained to make the best prediction by matching the predictions of the pseudo-label\footnote{A label that is generated by a model's prediction rather than being manually assigned by a human annotator.} with the ones generated under the strongly augmented image \citep{sohn2020fixmatch,slijepcevic2022radio}. \citet{slijepcevic2022radio} used Tang network classifier, in an SSL manner. They used MiraBest data (labeled) and the Radio Galaxy Zoo data release 1 (unlabelled). It was shown that the SSL strategy was able to extract knowledge from the unlabelled data thus achieving higher accuracy compared to the Tang classifier of the MiraBest data (baseline).

\subsubsection{N-shot learning}
N-shot learning algorithms are designed to leverage limited supervised information available (labeled data set) to make accurate predictions while avoiding overfitting challenges. Types of N-shot learning include Few-Shot Learning (FSL), One-Shot Learning (OSL), and Zero-Shot Learning (ZSL). \citet{samudre2022data} applied an FSL approach based on a Siamese neural network \citep{koch2015siamese}$^*$. The twin network model achieved an average precision of 74.2\%, a recall of 74.0\%, and an F$_{1}$ of 74.1\% for the classification of compact, FRI, FRII, and Bent radio galaxies \citep{samudre2022data}. In their experimentation, a sample size of 2708 radio galaxies was used. The samples were composed of selections from FRICAT, FRIICAT, CoNFIG, and Proctor data catalogs. While this approach has shown excellent performance on standard benchmark data sets, the twin network was found to yield relatively lower performance compared to the state-of-the-art supervised machine learning approaches on real data sets.   
%\subsubsection{contrastive learning methods???}

\subsection{Beyond classification}

\subsubsection{Anomaly detection}
In the context of astronomy, anomalies can be defined as  undiscovered and serendipitous astrophysical objects and phenomena  \citep{10.1093/mnras/sty3461,Lochner_2021} - peculiar objects having unexpected properties. With large data sets generated by radio telescopes, such as the EMU generating $\sim$70 million radio sources \citep{2011PASA...28..215N}, the SKA1 All-Sky continuum survey (SASS1), which is expected to generate $\sim$500 million radio sources, or the SKA2 All-Sky continuum Survey (SASS2), which is expected to increase to $\sim$3500 million radio sources \citep{https://doi.org/10.48550/arxiv.1412.6076}, the odds of discovering unknown unique objects are beyond doubt. Machine learning continues to play a critical role in unlocking discoveries by unpacking deep patterns in massive data sets. Hence, such automatic process supplements manual inspection of the objects to annotate new interesting radio sources and separate them from artifacts and already known sources.

Anomaly detection is mainly an unsupervised task where no labelled data is required.  In radio astronomy, there are few anomaly detection applications that can be referenced. \cite{Polsterer_2016} and \cite{Mostert_2021} investigated self-organizing maps to identify categories of radio galaxies using the Radio Galaxy Zoo Citizen project and LoTSS data, respectively. The identified objects that did not fall in any category of the known galaxies were annotated as outliers. In addition, \cite{Lochner_2021} developed an active anomaly detection algorithm\footnote{Active anomaly detection is an anomaly detection approach based on active learning. Active learning involves leveraging the expertise of a domain expert and the computational power of machine learning to improve the efficiency and effectiveness of the learning process.} that uses isolation forest and local outlier factor algorithms. In their paper, the anomaly detector is coupled with user feedback (based on interest). The algorithm detects and flags outliers and the user scores the results, which are then used to suppress dissimilar objects and display similar ones. 

Anomaly detection is mainly challenging because some identified anomalies may be artifacts introduced during data recording, calibration, and reduction procedures. Further to \cite{Lochner_2021}, some flagged anomalies may not be of interest to the research objectives of the astronomer. Therefore, the identified anomalies largely depend on the focus area of the astronomer and hence the relevance of the anomalies to a study may not be easily captured by machine/deep learning algorithms. Despite the progress achieved in the exploitation of machine intelligence, anomaly detection remains to be a challenging field of research.

\subsubsection{Source extraction}
Automated source finding and parameterization are necessary for next-generation radio interferometric surveys to extract radio sources, as these sources often lack clear boundaries and exhibit luminosity decay/diffuse from the center, making it challenging to distinguish them from noise in an image.

The development of deep learning-based techniques has been on the rise to solve the challenge of extracting compact and diffuse sources alike. Application of different architectural designs and implementations of CNNs have been explored, such as the simple CNN in ConvoSource \citep{https://doi.org/10.48550/arxiv.1910.03631}, Mask R-CNN \citep{he2017mask} in Astro R-CNN, and Tiramisu \citep{10.1007/978-3-030-89691-1_38} - recent semantic segmentation based on U-Net \citep{10.1007/978-3-319-24574-4_28}. These methods have shown that the use of deep learning methodologies in automatic detection and extraction of radio sources is robust and achieves high accuracies of above 90\%. In addition, they have shown significant improvements in classifying extended sources, for instance, the Tiramisu semantic segmentation by \citet{10.1007/978-3-030-89691-1_38} achieves an accuracy of 97\% though with a small sample size of 2,348 sources (where 320 sources are extended).

In essence, the latest state-of-art deep learning methodologies are promising alternatives to the dominant tools like PyBDSF. However, the deep learning algorithms' performance is found to be limiting when the images are noisy, the sources are faint or have diffuse morphological structure.

\section{Opportunities, challenges, and outlook}
\label{sectn:discussion}

Computer intelligence is having a remarkable impact on radio astronomy. A plethora of new insightful scientific work is published every year, resulting in even better and more accurate models that generalize well. As a result, there are now open opportunities to develop robust models that are capable of generating predictions across surveys from different yet related next-generation telescopes (such as LOFAR, MeerKAT, and SKA). Furthermore, these models would require slight to no modification once a new data release is made available. This highlights the potential for further scientific progress in utilizing raw radio image cubes generated by modern telescopes, through the incorporation of computer intelligence.

Despite the predominance of massive high-resolution data sets from modern telescopes, there is limited availability of annotated data sets. As a result, this hinders the ability to fully utilize and exploit the potential of artificial intelligence in the data-rich field. While there are developed strategies (such as data augmentation, semi-supervised learning and weakly supervised approaches) leveraging small data samples \citep{2019MNRAS.488.3358T,slijepcevic2022radio}, such strategies cannot match the diverse and unique astrophysical phenomena embedded in the massive radio images. Therefore, this calls for continued collaborative efforts in the generation of annotated machine/deep learning-ready data sets while considering compute resources.

Radio astronomy is a data-rich and compute-intensive field, hence exploitation of scalable platforms and software is paramount. In order to train a model using techniques such as SOM \citep{Galvin_2019}, SVM \citep{2021AJ....161...94S} and DCNN \citep{2021AJ....161...94S}, a significant amount of computing resources are required. For instance, DCNNs typically require large amounts of images in order to learn over a million parameters that characterize a model. Therefore, as the available data in astronomy increases exponentially, and more specialized machine/deep learning algorithms are developed, the demand for highly scalable computing performance is inevitable. High-performance computing (HPC), graphical processing units (GPUs) and distributed computing are often used to run such algorithms. In particular, big data (radio astronomical data) requires sophisticated methodologies to efficiently query and process large volumes of data. Despite the availability of numerous studies, as discussed in this review paper, there is still a wide gap in the utilization of scalable pipelines that allow for more efficient parallel and distributed machine/deep learning computations. Pipelines that would take advantage of some of the storage formats of the radio astronomical survey data. For instance, LOFAR uses H5parm,  a Hierarchical Data Format version 5 (HDF5) compliant file format, which provides an excellent basis for applying Apache Spark\footnote{https://spark.apache.org/}, a Big data processing ecosystem. 

Indexing of identified radio sources is a prerequisite for fast retrieval of radio galaxies of similar/dissimilar morphological attributes. However, as this topic is hardly addressed in the literature covered, it highlights the existing research gap in radio astronomy that needs to be filled. Image indexing and/or retrieval is the process of finding objects (images) that have similar characteristics with varied shapes and sizes. Having developed a database of known and unknown (anomalous) radio astronomical structures, it is of great importance to develop a system that would aid in the quick retrieval of galaxies with similar morphological characteristics \citep{AbdElAziz2017AutomaticDO}. Ideally, identified objects are indexed with a hashing function that minimizes the distances between perceptually similar objects and maximizes those of dissimilar objects. This is a paradigm that has seen a lot of progress in recent years with the development of deep hashing methods \citep{luo2020survey}, a paradigm that to our knowledge is yet to be leveraged in radio astronomy.
\vfill
\clearpage
\section{Conclusion}
\label{sectn:conclusion}

Radio astronomy is in the era of Big Data, presenting ubiquitous opportunities that necessitate extensive automation of data processing, exploration, and scientific exploitation. This will unravel the cosmology space, if modern telescopes reach their scientific goals. In this regard, astronomers have taken undue advantage of the deep neural network revolution in computer vision with notable success. 

In this survey paper, we have presented a detailed literature overview of the data and algorithmic advances in data curation pipelines, data preprocessing strategies, and cutting-edge machine intelligence methods. New scientific works that involve the development of robust and accurate novel models have emerged in the field of radio astronomy. These models can capture the diverse and unique astrophysical phenomena found in large radio images through the use of techniques like data augmentation, semi-supervised learning, and weakly supervised approaches. This has opened up the possibility of creating models that can accurately predict the outcomes of surveys conducted with telescopes like LOFAR and SKA, without significant modification when new data becomes available.

The survey revealed that there has been little exploration of image indexing and retrieval within the field of radio astronomy, even though it is an essential step for quickly retrieving radio images with similar or dissimilar morphological structures. This area of research offers considerable potential for future investigation.

\section{Appendix}
%\vfill

\begin{figure}[h!]
\centering
\footnotesize
\includegraphics{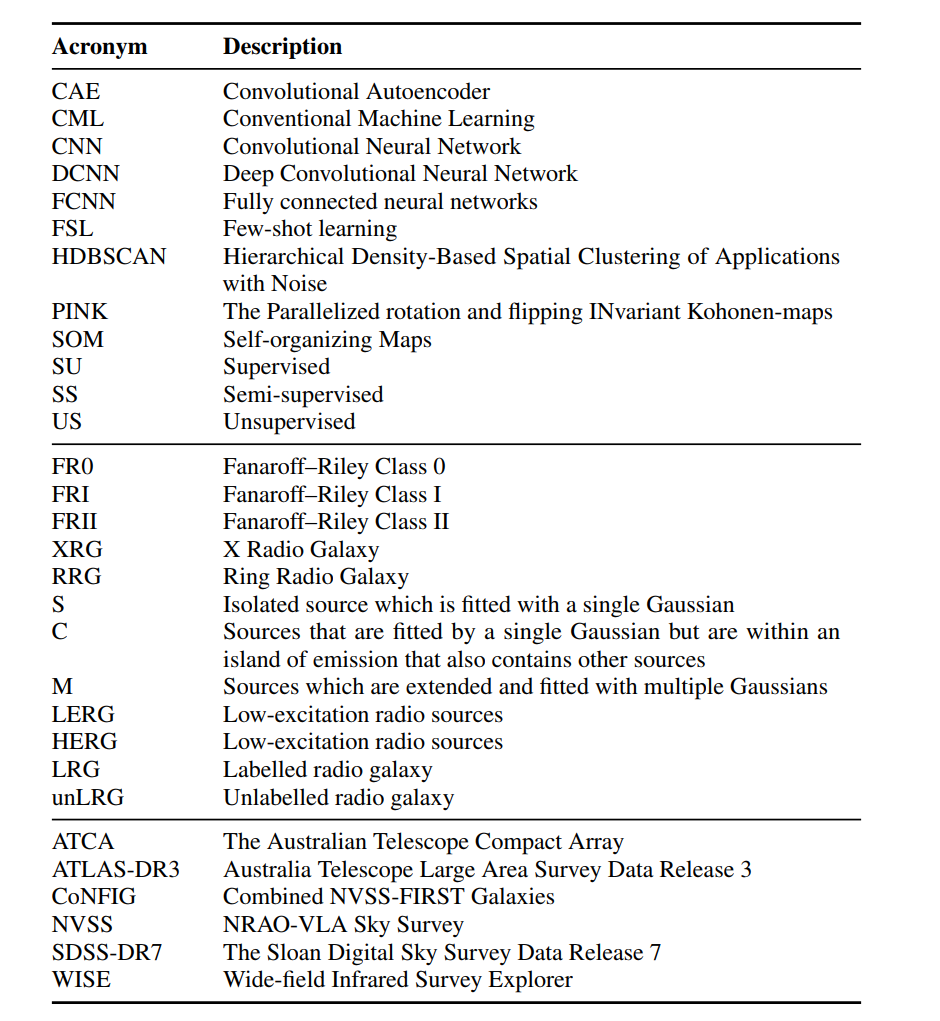}
\caption{The abbreviations are categorized in three sections, with the top section representing algorithm keywords, the middle section representing galaxies, and the bottom section representing astronomical surveys.}
\label{tab:abbreviations}
\end{figure}

\newpage
\bibliographystyle{unsrtnat}
%\bibliographystyle{abbrvnat}
% \bibliographystyle{plain}
%\bibliography{references}  %%% Uncomment this line and comment out the ``thebibliography'' section below to use the external .bib file (using bibtex) .

%%% Uncomment this section and comment out the \bibliography{references} line above to use inline references.

\begin{thebibliography}{1}
\ifx \showCODEN    \undefined \def \showCODEN     #1{\unskip}     \fi
\ifx \showDOI      \undefined \def \showDOI       #1{#1}\fi
\ifx \showISBNx    \undefined \def \showISBNx     #1{\unskip}     \fi
\ifx \showISBNxiii \undefined \def \showISBNxiii  #1{\unskip}     \fi
\ifx \showISSN     \undefined \def \showISSN      #1{\unskip}     \fi
\ifx \showLCCN     \undefined \def \showLCCN      #1{\unskip}     \fi
\ifx \shownote     \undefined \def \shownote      #1{#1}          \fi
\ifx \showarticletitle \undefined \def \showarticletitle #1{#1}   \fi
\ifx \showURL      \undefined \def \showURL       {\relax}        \fi
% The following commands are used for tagged output and should be
% invisible to TeX
\providecommand\bibfield[2]{#2}
\providecommand\bibinfo[2]{#2}
\providecommand\natexlab[1]{#1}
\providecommand\showeprint[2][]{arXiv:#2}

\bibitem[{Farnes et al.}(2018)]{galaxies6040120}
{Farnes et al.}
\newblock Science pipelines for the square kilometre array.
\newblock \emph{Galaxies}, 6\penalty0 (4), 2018.
\newblock ISSN 2075-4434.
\newblock \doi{10.3390/galaxies6040120}.

\bibitem[{Labate et al.}(2022)]{10.1117/1.JATIS.8.1.011024}
{Labate et al.}
\newblock {Highlights of the Square Kilometre Array Low Frequency (SKA-LOW)
  Telescope}.
\newblock \emph{Journal of Astronomical Telescopes, Instruments, and Systems},
  8\penalty0 (1):\penalty0 011024, 2022.
\newblock \doi{10.1117/1.JATIS.8.1.011024}.

\bibitem[Swart et~al.(2022)Swart, Dewdney, and
  Cremonini]{10.1117/1.JATIS.8.1.011021}
Gerhard~P. Swart, Peter~E. Dewdney, and Andrea Cremonini.
\newblock {Highlights of the SKA1-Mid telescope architecture}.
\newblock \emph{Journal of Astronomical Telescopes, Instruments, and Systems},
  8\penalty0 (1):\penalty0 011021, 2022.
\newblock \doi{10.1117/1.JATIS.8.1.011021}.

\bibitem[Booth and Jonas(2012)]{booth2012overview}
RS~Booth and JL~Jonas.
\newblock {An overview of the MeerKAT project}.
\newblock \emph{African Skies}, 16:\penalty0 101, 2012.

\bibitem[{Lonsdale et al.}(2009)]{lonsdale2009murchison}
{Lonsdale et al.}
\newblock {The murchison widefield array: Design overview}.
\newblock \emph{Proceedings of the IEEE}, 97\penalty0 (8):\penalty0 1497--1506,
  2009.

\bibitem[{Haarlem et al.}(2013)]{refId0}
{Haarlem et al.}
\newblock Lofar: The low-frequency array.
\newblock \emph{Astronomy \& Astrophysics}, 556:\penalty0 A2, 2013.
\newblock \doi{10.1051/0004-6361/201220873}.

\bibitem[An(2019)]{An2019ScienceOA}
Tao An.
\newblock {Science opportunities and challenges associated with SKA big data}.
\newblock \emph{Science China Physics, Mechanics \& Astronomy}, 62:\penalty0
  1--6, 2019.

\bibitem[{Norris et al.}(2011)]{2011PASA...28..215N}
{Norris et al.}
\newblock {EMU: evolutionary map of the universe}.
\newblock \emph{Publications of the Astronomical Society of Australia},
  28\penalty0 (3):\penalty0 215--248, 2011.

\bibitem[{Norris et al.}(2014)]{https://doi.org/10.48550/arxiv.1412.6076}
{Norris et al.}
\newblock The ska mid-frequency all-sky continuum survey: Discovering the
  unexpected and transforming radio-astronomy, 2014.

\bibitem[Ray(2016)]{ray2016discovering}
PN~Ray.
\newblock {Discovering the Unexpected in Astronomical Survey Data}.
\newblock \emph{Publ Astron Soc Aust}, 34\penalty0 (10), 2016.

\bibitem[{Burke et al.}(2019)]{burke2019introduction}
{Burke et al.}
\newblock \emph{{An introduction to radio astronomy}}.
\newblock Cambridge University Press, 2019.

\bibitem[{Hu et al.}(2020)]{10.1093/mnras/staa3087}
{Hu et al.}
\newblock {Telescope performance real-time monitoring based on machine
  learning}.
\newblock \emph{Monthly Notices of the Royal Astronomical Society},
  500\penalty0 (1):\penalty0 388--396, 10 2020.
\newblock ISSN 0035-8711.
\newblock \doi{10.1093/mnras/staa3087}.

\bibitem[{Mesarcik et al.}(2020)]{10.1093/mnras/staa1412}
{Mesarcik et al.}
\newblock {Deep learning assisted data inspection for radio astronomy}.
\newblock \emph{Monthly Notices of the Royal Astronomical Society},
  496\penalty0 (2):\penalty0 1517--1529, 05 2020.
\newblock ISSN 0035-8711.
\newblock \doi{10.1093/mnras/staa1412}.

\bibitem[Yatawatta and Avruch(2021)]{Yatawatta_2021}
Sarod Yatawatta and Ian~M Avruch.
\newblock Deep reinforcement learning for smart calibration of radio
  telescopes.
\newblock \emph{Monthly Notices of the Royal Astronomical Society},
  505\penalty0 (2):\penalty0 2141--2150, may 2021.
\newblock \doi{10.1093/mnras/stab1401}.

\bibitem[{Sun et al.}(2022)]{10.1093/mnras/stac570}
{Sun et al.}
\newblock {A robust RFI identification for radio interferometry based on a
  convolutional neural network}.
\newblock \emph{Monthly Notices of the Royal Astronomical Society},
  512\penalty0 (2):\penalty0 2025--2033, 03 2022.
\newblock ISSN 0035-8711.
\newblock \doi{10.1093/mnras/stac570}.

\bibitem[{Wijnholds et al.}(2010)]{5355494}
{Wijnholds et al.}
\newblock {Calibration challenges for future radio telescopes}.
\newblock \emph{IEEE Signal Processing Magazine}, 27\penalty0 (1):\penalty0
  30--42, 2010.
\newblock \doi{10.1109/MSP.2009.934853}.

\bibitem[{Lukic et
  al.}(2019{\natexlab{a}})]{https://doi.org/10.48550/arxiv.1910.03631}
{Lukic et al.}
\newblock {ConvoSource: radio-astronomical source-finding with convolutional
  neural networks}.
\newblock \emph{Galaxies}, 8\penalty0 (1):\penalty0 3, 2019{\natexlab{a}}.

\bibitem[{Pino et al.}(2021)]{10.1007/978-3-030-89691-1_38}
{Pino et al.}
\newblock Semantic segmentation of radio-astronomical images.
\newblock In Yanio Hern{\'a}ndez~Heredia, Vladimir Mili{\'a}n~N{\'u}{\~{n}}ez,
  and Jos{\'e} Ruiz~Shulcloper, editors, \emph{Progress in Artificial
  Intelligence and Pattern Recognition}, pages 393--403, Cham, 2021. Springer
  International Publishing.
\newblock ISBN 978-3-030-89691-1.

\bibitem[{Lukic et al.}(2018)]{lukic2018radio}
{Lukic et al.}
\newblock {Radio Galaxy Zoo: compact and extended radio source classification
  with deep learning}.
\newblock \emph{Monthly Notices of the Royal Astronomical Society},
  476\penalty0 (1):\penalty0 246--260, 2018.

\bibitem[{Wu et al.}(2018)]{10.1093/mnras/sty2646}
{Wu et al.}
\newblock {Radio Galaxy Zoo: Claran – a deep learning classifier for radio
  morphologies}.
\newblock \emph{Monthly Notices of the Royal Astronomical Society},
  482\penalty0 (1):\penalty0 1211--1230, 10 2018.
\newblock ISSN 0035-8711.
\newblock \doi{10.1093/mnras/sty2646}.

\bibitem[{Mostert et al.}(2021)]{Mostert_2021}
{Mostert et al.}
\newblock {Unveiling the rarest morphologies of the LOFAR Two-metre Sky Survey
  radio source population with self-organised maps}.
\newblock \emph{Astronomy \& Astrophysics}, 645:\penalty0 A89, 2021.

\bibitem[Aziz et~al.(2017)Aziz, Selim, and Xiong]{AbdElAziz2017AutomaticDO}
Mohamed Abd~El Aziz, Ibrahim Selim, and Shengwu Xiong.
\newblock {Automatic Detection of Galaxy Type From Datasets of Galaxies Image
  Based on Image Retrieval Approach}.
\newblock \emph{Scientific Reports}, 7, 2017.

\bibitem[{Shimwell et al.}(2022{\natexlab{a}})]{Shimwell_2022}
{Shimwell et al.}
\newblock The {LOFAR} two-metre sky survey.
\newblock \emph{Astronomy \& Astrophysics}, 659:\penalty0 A1, feb
  2022{\natexlab{a}}.
\newblock \doi{10.1051/0004-6361/202142484}.

\bibitem[Fanaroff and Riley(1974)]{fanaroff1974morphology}
Bernard~L Fanaroff and Julia~M Riley.
\newblock {The morphology of extragalactic radio sources of high and low
  luminosity}.
\newblock \emph{Monthly Notices of the Royal Astronomical Society},
  167\penalty0 (1):\penalty0 31P--36P, 1974.

\bibitem[Rudnick and Owen(1976)]{rudnick1976head}
Lawrence Rudnick and Frazer~N. Owen.
\newblock {Head-tail radio sources in clusters of galaxies}.
\newblock \emph{The Astrophysical Journal}, 203:\penalty0 L107--L111, 1976.

\bibitem[Baldi et~al.(2015)Baldi, Capetti, and Giovannini]{Baldi2015APS}
Ranieri~D Baldi, Alessandro Capetti, and Gabriele Giovannini.
\newblock {Pilot study of the radio-emitting AGN population: the emerging new
  class of FR 0 radio-galaxies}.
\newblock \emph{Astronomy \& Astrophysics}, 576:\penalty0 A38, 2015.

\bibitem[Baldi et~al.(2018)Baldi, Capetti, and Massaro]{baldi2018fr0cat}
RD~Baldi, Alessandro Capetti, and F~Massaro.
\newblock {FR0CAT: a FIRST catalog of FR 0 radio galaxies}.
\newblock \emph{Astronomy \& Astrophysics}, 609:\penalty0 A1, 2018.

\bibitem[Proctor(2011)]{proctor2011morphological}
DD~Proctor.
\newblock {Morphological annotations for groups in the first database}.
\newblock \emph{The Astrophysical Journal Supplement Series}, 194\penalty0
  (2):\penalty0 31, 2011.

\bibitem[{Wilkinson et al.}(2016)]{wilkinson2016a}
{Wilkinson et al.}
\newblock The fair guiding principles for scientific data management and
  stewardship.
\newblock \emph{Scientific Data}, 3:\penalty0 160018, 2016.

\bibitem[{Mireille et al.}(2022)]{https://doi.org/10.48550/arxiv.2012.09273}
{Mireille et al.}
\newblock {Radio Astronomy Visibility Data Discovery and Access Using IVOA
  Standards}.
\newblock In \emph{{Astronomical Society of the Pacific Conference Series}},
  volume 532, page 443, 2022.

\bibitem[O'Toole and Tocknell(2022)]{https://doi.org/10.48550/arxiv.2203.10710}
Simon O'Toole and James Tocknell.
\newblock {FAIR standards for astronomical data}.
\newblock \emph{arXiv preprint arXiv:2203.10710}, 2022.

\bibitem[{Mohan} and {Rafferty}(2015)]{2015ascl.soft02007M}
Niruj {Mohan} and David {Rafferty}.
\newblock {Pybdsf: Python blob detection and source finder}.
\newblock \emph{Astrophysics Source Code Library}, pages ascl--1502, 2015.

\bibitem[{Hales et al.}(2012)]{10.1111/j.1365-2966.2012.21373.x}
{Hales et al.}
\newblock {BLOBCAT: software to catalogue flood-filled blobs in radio images of
  total intensity and linear polarization}.
\newblock \emph{Monthly Notices of the Royal Astronomical Society},
  425\penalty0 (2):\penalty0 979--996, 09 2012.
\newblock ISSN 0035-8711.
\newblock \doi{10.1111/j.1365-2966.2012.21373.x}.

\bibitem[{Hancock et al.}(2012)]{10.1111/j.1365-2966.2012.20768.x}
{Hancock et al.}
\newblock {Compact continuum source finding for next generation radio surveys}.
\newblock \emph{Monthly Notices of the Royal Astronomical Society},
  422\penalty0 (2):\penalty0 1812--1824, 04 2012.
\newblock ISSN 0035-8711.
\newblock \doi{10.1111/j.1365-2966.2012.20768.x}.

\bibitem[{Hopkins et al.}(2015)]{Hopkins_2015}
{Hopkins et al.}
\newblock The {ASKAP}/{EMU} source finding data challenge.
\newblock \emph{Publications of the Astronomical Society of Australia}, 32,
  2015.
\newblock \doi{10.1017/pasa.2015.37}.

\bibitem[{Pence et al.}(2010)]{Pence2010DefinitionOT}
{Pence et al.}
\newblock {Definition of the Flexible Image Transport System (FITS), version
  3.0}.
\newblock \emph{Astronomy and Astrophysics}, 524, 2010.

\bibitem[{Smith et al.}(2014)]{2014ascl.soft11023W}
{Smith et al.}
\newblock {NDF: Extensible N-dimensional Data Format Library}.
\newblock \emph{Astrophysics Source Code Library}, pages ascl--1411, 2014.

\bibitem[{van Diepen}(2015)]{VANDIEPEN2015174}
G.N.J. {van Diepen}.
\newblock {Casacore Table Data System and its use in the MeasurementSet}.
\newblock \emph{Astronomy and Computing}, 12:\penalty0 174--180, 2015.
\newblock ISSN 2213-1337.
\newblock \doi{https://doi.org/10.1016/j.ascom.2015.06.002}.

\bibitem[Greisen(2011)]{greisen2011fits}
Eric~W Greisen.
\newblock {The FITS interferometry data interchange convention—Revised}.
\newblock \emph{{AIPS Memo Series, 114r, Socorro, New Mexico, USA}}, 2011.

\bibitem[Greisen(2012)]{greisen2012aips}
Eric~W Greisen.
\newblock {AIPS FITS file format}.
\newblock \emph{AIPS Memo}, 117, 2012.

\bibitem[Price et~al.(2014)Price, Barsdell, and
  Greenhill]{https://doi.org/10.48550/arxiv.1411.0507}
Danny~C. Price, Benjamin~R. Barsdell, and Lincoln~J. Greenhill.
\newblock Is hdf5 a good format to replace uvfits?, 2014.

\bibitem[Miraghaei and Best(2017)]{10.1093/mnras/stx007}
H.~Miraghaei and P.~N. Best.
\newblock {The nuclear properties and extended morphologies of powerful radio
  galaxies: the roles of host galaxy and environment}.
\newblock \emph{Monthly Notices of the Royal Astronomical Society},
  466\penalty0 (4):\penalty0 4346--4363, 01 2017.
\newblock ISSN 0035-8711.
\newblock \doi{10.1093/mnras/stx007}.

\bibitem[Gendre et~al.(2010)Gendre, Best, and
  Wall]{10.1111/j.1365-2966.2010.16413.x}
M.~A. Gendre, P.~N. Best, and J.~V. Wall.
\newblock {The Combined NVSS–FIRST Galaxies (CoNFIG) sample – II.
  Comparison of space densities in the Fanaroff–Riley dichotomy}.
\newblock \emph{Monthly Notices of the Royal Astronomical Society},
  404\penalty0 (4):\penalty0 1719--1732, 05 2010.
\newblock ISSN 0035-8711.
\newblock \doi{10.1111/j.1365-2966.2010.16413.x}.

\bibitem[Aniyan and Thorat(2017)]{aniyan2017classifying}
AK~Aniyan and Kshitij Thorat.
\newblock {Classifying radio galaxies with the convolutional neural network}.
\newblock \emph{The Astrophysical Journal Supplement Series}, 230\penalty0
  (2):\penalty0 20, 2017.

\bibitem[{Ma et al.}(2019{\natexlab{a}})]{ma2019machine}
{Ma et al.}
\newblock {A machine learning based morphological classification of 14,245
  radio agns selected from the best--heckman sample}.
\newblock \emph{The Astrophysical Journal Supplement Series}, 240\penalty0
  (2):\penalty0 34, 2019{\natexlab{a}}.

\bibitem[Capetti et~al.(2017{\natexlab{a}})Capetti, Massaro, and
  Baldi]{capetti2017fricat}
Alessandro Capetti, Francesco Massaro, and Ranieri~Diego Baldi.
\newblock {FRICAT: a FIRST catalog of FR I radio galaxies}.
\newblock \emph{Astronomy \& Astrophysics}, 598:\penalty0 A49,
  2017{\natexlab{a}}.

\bibitem[Capetti et~al.(2017{\natexlab{b}})Capetti, Massaro, and
  Baldi]{capetti2017friicat}
A.~Capetti, Francesco Massaro, and Ranieri Baldi.
\newblock {FRIICAT: A FIRST catalog of FR II radio galaxies}.
\newblock \emph{Astronomy \& Astrophysics}, 601, 02 2017{\natexlab{b}}.
\newblock \doi{10.1051/0004-6361/201630247}.

\bibitem[Best and Heckman(2012)]{10.1111/j.1365-2966.2012.20414.x}
P.~N. Best and T.~M. Heckman.
\newblock {On the fundamental dichotomy in the local radio-AGN population:
  accretion, evolution and host galaxy properties}.
\newblock \emph{Monthly Notices of the Royal Astronomical Society},
  421\penalty0 (2):\penalty0 1569--1582, 03 2012.
\newblock ISSN 0035-8711.
\newblock \doi{10.1111/j.1365-2966.2012.20414.x}.

\bibitem[Gendre and Wall(2008)]{gendre2008combined}
MA~Gendre and JV~Wall.
\newblock {The Combined NVSS--FIRST Galaxies (CoNFIG) sample--I. Sample
  definition, classification and evolution}.
\newblock \emph{Monthly Notices of the Royal Astronomical Society},
  390\penalty0 (2):\penalty0 819--828, 2008.

\bibitem[{Shimwell et al.}(2019)]{shimwell2019lofar}
{Shimwell et al.}
\newblock {The LOFAR two-metre sky survey-II. First data release}.
\newblock \emph{Astronomy \& Astrophysics}, 622:\penalty0 A1, 2019.

\bibitem[{Shimwell et al.}(2022{\natexlab{b}})]{shimwell2022lofar}
{Shimwell et al.}
\newblock {The LOFAR Two-metre Sky Survey-V. Second data release}.
\newblock \emph{Astronomy \& Astrophysics}, 659:\penalty0 A1,
  2022{\natexlab{b}}.

\bibitem[Ntwaetsile and Geach(2021)]{10.1093/mnras/stab271}
Kushatha Ntwaetsile and James~E Geach.
\newblock {Rapid sorting of radio galaxy morphology using Haralick features}.
\newblock \emph{Monthly Notices of the Royal Astronomical Society},
  502\penalty0 (3):\penalty0 3417--3425, 02 2021.
\newblock ISSN 0035-8711.
\newblock \doi{10.1093/mnras/stab271}.

\bibitem[{Mingo et al.}(2019)]{mingo2019revisiting}
{Mingo et al.}
\newblock {Revisiting the Fanaroff--Riley dichotomy and radio-galaxy morphology
  with the LOFAR Two-Metre Sky Survey (LoTSS)}.
\newblock \emph{Monthly Notices of the Royal Astronomical Society},
  488\penalty0 (2):\penalty0 2701--2721, 2019.

\bibitem[{Lukic et al.}(2019{\natexlab{b}})]{Lukic_2019}
{Lukic et al.}
\newblock Morphological classification of radio galaxies: capsule networks
  versus convolutional neural networks.
\newblock \emph{Monthly Notices of the Royal Astronomical Society},
  487\penalty0 (2):\penalty0 1729--1744, may 2019{\natexlab{b}}.
\newblock \doi{10.1093/mnras/stz1289}.

\bibitem[Cheung(2007)]{cheung2007first}
CC~Cheung.
\newblock {FIRST “Winged” And X-Shaped radio source candidates}.
\newblock \emph{The Astronomical Journal}, 133\penalty0 (5):\penalty0 2097,
  2007.

\bibitem[{Becker et al.}(2021)]{becker2021cnn}
{Becker et al.}
\newblock {CNN architecture comparison for radio galaxy classification}.
\newblock \emph{Monthly Notices of the Royal Astronomical Society},
  503\penalty0 (2):\penalty0 1828--1846, 2021.

\bibitem[Alhassan et~al.(2018)Alhassan, Taylor, and Vaccari]{alhassan2018first}
Wathela Alhassan, AR~Taylor, and Mattia Vaccari.
\newblock {The FIRST Classifier: compact and extended radio galaxy
  classification using deep Convolutional Neural Networks}.
\newblock \emph{Monthly Notices of the Royal Astronomical Society},
  480\penalty0 (2):\penalty0 2085--2093, 2018.

\bibitem[{Kummer et al.}(2022)]{kummer2022radio}
{Kummer et al.}
\newblock {Radio Galaxy Classification with wGAN-Supported Augmentation}.
\newblock \emph{arXiv preprint arXiv:2206.15131}, 2022.

\bibitem[Scaife and Porter(2021)]{scaife2021fanaroff}
Anna~MM Scaife and Fiona Porter.
\newblock {Fanaroff--Riley classification of radio galaxies using
  group-equivariant convolutional neural networks}.
\newblock \emph{Monthly Notices of the Royal Astronomical Society},
  503\penalty0 (2):\penalty0 2369--2379, 2021.

\bibitem[Sadeghi et~al.(2021)Sadeghi, Javaherian, and
  Miraghaei]{2021AJ....161...94S}
Mohammad Sadeghi, Mohsen Javaherian, and Halime Miraghaei.
\newblock {Morphological-based Classifications of Radio Galaxies Using
  Supervised Machine-learning Methods Associated with Image Moments}.
\newblock \emph{The Astronomical Journal}, 161\penalty0 (2):\penalty0 94, 2021.

\bibitem[{Slijepcevic et al.}(2022)]{slijepcevic2022radio}
{Slijepcevic et al.}
\newblock {Radio Galaxy Zoo: using semi-supervised learning to leverage large
  unlabelled data sets for radio galaxy classification under data set shift}.
\newblock \emph{Monthly Notices of the Royal Astronomical Society},
  514\penalty0 (2):\penalty0 2599--2613, 2022.

\bibitem[{Samudre et al.}(2022)]{samudre2022data}
{Samudre et al.}
\newblock {Data-efficient classification of radio galaxies}.
\newblock \emph{Monthly Notices of the Royal Astronomical Society},
  509\penalty0 (2):\penalty0 2269--2280, 2022.

\bibitem[Maslej-Kre{\v{s}}{\v{n}}{\'a}kov{\'a}
  et~al.(2021)Maslej-Kre{\v{s}}{\v{n}}{\'a}kov{\'a}, El~Bouchefry, and
  Butka]{maslej2021morphological}
Viera Maslej-Kre{\v{s}}{\v{n}}{\'a}kov{\'a}, Khadija El~Bouchefry, and Peter
  Butka.
\newblock {Morphological classification of compact and extended radio galaxies
  using convolutional neural networks and data augmentation techniques}.
\newblock \emph{Monthly Notices of the Royal Astronomical Society},
  505\penalty0 (1):\penalty0 1464--1475, 2021.

\bibitem[{Banfield et al.}(2015)]{banfield2015radio}
{Banfield et al.}
\newblock {Radio Galaxy Zoo: host galaxies and radio morphologies derived from
  visual inspection}.
\newblock \emph{Monthly Notices of the Royal Astronomical Society},
  453\penalty0 (3):\penalty0 2326--2340, 2015.

\bibitem[{Tang et al.}(2022)]{tang2022radio}
{Tang et al.}
\newblock {Radio Galaxy Zoo: giant radio galaxy classification using
  multidomain deep learning}.
\newblock \emph{Monthly Notices of the Royal Astronomical Society},
  510\penalty0 (3):\penalty0 4504--4524, 2022.

\bibitem[{Ralph et al.}(2019)]{ralph2019radio}
{Ralph et al.}
\newblock {Radio galaxy zoo: Unsupervised clustering of convolutionally
  auto-encoded radio-astronomical images}.
\newblock \emph{Publications of the Astronomical Society of the Pacific},
  131\penalty0 (1004):\penalty0 108011, 2019.

\bibitem[{Ma et al.}(2019{\natexlab{b}})]{ma2019classification}
{Ma et al.}
\newblock {Classification of radio galaxy images with semi-supervised
  learning}.
\newblock In \emph{{International Conference on Data Mining and Big Data}},
  pages 191--200. Springer, 2019{\natexlab{b}}.

\bibitem[Wee and Banister(2016)]{wee2016write}
Bert~Van Wee and David Banister.
\newblock {How to write a literature review paper?}
\newblock \emph{Transport reviews}, 36\penalty0 (2):\penalty0 278--288, 2016.

\bibitem[Wohlin(2014)]{Wohlin2014GuidelinesFS}
Claes Wohlin.
\newblock {Guidelines for snowballing in systematic literature studies and a
  replication in software engineering}.
\newblock In \emph{{EASE '14}}, 2014.

\bibitem[Fluke and Jacobs(2020)]{fluke2020surveying}
Christopher~J Fluke and Colin Jacobs.
\newblock {Surveying the reach and maturity of machine learning and artificial
  intelligence in astronomy}.
\newblock \emph{Wiley Interdisciplinary Reviews: Data Mining and Knowledge
  Discovery}, 10\penalty0 (2):\penalty0 e1349, 2020.

\bibitem[{Wang et al.}(2018)]{wang2018computational}
{Wang et al.}
\newblock {Computational intelligence in astronomy: A survey}.
\newblock \emph{International Journal of Computational Intelligence Systems},
  11\penalty0 (1):\penalty0 575, 2018.

\bibitem[{Bowles et al.}(2021)]{bowles2021attention}
{Bowles et al.}
\newblock {Attention-gating for improved radio galaxy classification}.
\newblock \emph{Monthly Notices of the Royal Astronomical Society},
  501\penalty0 (3):\penalty0 4579--4595, 2021.

\bibitem[Becker and Grobler(2019)]{9015881}
Burger Becker and Trienko Grobler.
\newblock {Classification of Fanaroff-Riley Radio Galaxies using Conventional
  Machine Learning Techniques}.
\newblock In \emph{{2019 International Multidisciplinary Information Technology
  and Engineering Conference (IMITEC)}}, pages 1--8, 2019.
\newblock \doi{10.1109/IMITEC45504.2019.9015881}.

\bibitem[{Galvin et al.}(2019)]{Galvin_2019}
{Galvin et al.}
\newblock Radio galaxy zoo: Knowledge transfer using rotationally invariant
  self-organizing maps.
\newblock \emph{Publications of the Astronomical Society of the Pacific},
  131\penalty0 (1004):\penalty0 108009, sep 2019.
\newblock \doi{10.1088/1538-3873/ab150b}.

\bibitem[Tang et~al.(2019)Tang, Scaife, and Leahy]{2019MNRAS.488.3358T}
Hongming Tang, Anna~MM Scaife, and JP~Leahy.
\newblock {Transfer learning for radio galaxy classification}.
\newblock \emph{Monthly Notices of the Royal Astronomical Society},
  488\penalty0 (3):\penalty0 3358--3375, 2019.

\bibitem[{Mostert et al.}(2022)]{mostert2022radio}
{Mostert et al.}
\newblock Radio source-component association for the lofar two-metre sky survey
  with region-based convolutional neural networks.
\newblock \emph{Astronomy \& Astrophysics}, 668, DEC 1 2022.
\newblock ISSN 0004-6361.
\newblock \doi{10.1051/0004-6361/202243478}.

\bibitem[{Rezaei et al.}(2022)]{rezaei2022decoras}
{Rezaei et al.}
\newblock {DECORAS: detection and characterization of radio-astronomical
  sources using deep learning}.
\newblock \emph{Monthly Notices of the Royal Astronomical Society},
  510\penalty0 (4):\penalty0 5891--5907, 2022.

\bibitem[{Vafaei et al.}(2019)]{vafaei2019deepsource}
{Vafaei et al.}
\newblock {DEEPSOURCE: point source detection using deep learning}.
\newblock \emph{Monthly Notices of the Royal Astronomical Society},
  484\penalty0 (2):\penalty0 2793--2806, 2019.

\bibitem[Gheller et~al.(2018)Gheller, Vazza, and Bonafede]{Gheller2018DeepLB}
Claudio Gheller, Franco Vazza, and Annalisa Bonafede.
\newblock {Deep learning based detection of cosmological diffuse radio
  sources}.
\newblock \emph{Monthly Notices of the Royal Astronomical Society}, 2018.

\bibitem[{Polsterer et al.}(2016)]{Polsterer_2016}
{Polsterer et al.}
\newblock {Parallelized rotation and flipping INvariant Kohonen maps (PINK) on
  GPUs}.
\newblock In \emph{{ESANN 2016: 24th European Symposium on Artificial Neural
  Networks, Computational Intelligence and Machine Learning. Bruges (Belgium),
  27-29 April 2016. Proceedings}}, pages 406--410. Bruges: i6doc. com, 2016.

\bibitem[Liu and Deng(2015)]{7486599}
Shuying Liu and Weihong Deng.
\newblock {Very deep convolutional neural network based image classification
  using small training sample size}.
\newblock In \emph{{2015 3rd IAPR Asian Conference on Pattern Recognition
  (ACPR)}}, pages 730--734, 2015.
\newblock \doi{10.1109/ACPR.2015.7486599}.

\bibitem[Peterson(2009)]{peterson2009k}
Leif~E Peterson.
\newblock {K-nearest neighbor}.
\newblock \emph{Scholarpedia}, 4\penalty0 (2):\penalty0 1883, 2009.

\bibitem[Cortes and Vapnik(1995)]{cortes1995support}
Corinna Cortes and Vladimir Vapnik.
\newblock {Support-vector networks}.
\newblock \emph{Machine learning}, 20\penalty0 (3):\penalty0 273--297, 1995.

\bibitem[{Ding et al.}(2021)]{DING202110121}
{Ding et al.}
\newblock Random radial basis function kernel-based support vector machine.
\newblock \emph{Journal of the Franklin Institute}, 358\penalty0 (18):\penalty0
  10121--10140, 2021.
\newblock ISSN 0016-0032.
\newblock \doi{https://doi.org/10.1016/j.jfranklin.2021.10.005}.

\bibitem[Banerjee et~al.(2013)Banerjee, Dunson, and
  Tokdar]{banerjee2013efficient}
Anjishnu Banerjee, David~B Dunson, and Surya~T Tokdar.
\newblock {Efficient Gaussian process regression for large datasets}.
\newblock \emph{Biometrika}, 100\penalty0 (1):\penalty0 75--89, 2013.

\bibitem[Freund and Schapire(1997)]{freund1997decision}
Yoav Freund and Robert~E Schapire.
\newblock {A decision-theoretic generalization of online learning and an
  application to boosting}.
\newblock \emph{Journal of computer and system sciences}, 55\penalty0
  (1):\penalty0 119--139, 1997.

\bibitem[Breiman(2001)]{breiman2001random}
Leo Breiman.
\newblock {Random forests}.
\newblock \emph{Machine learning}, 45\penalty0 (1):\penalty0 5--32, 2001.

\bibitem[Rish et~al.(2001)]{rish2001empirical}
Irina Rish et~al.
\newblock {An empirical study of the naive Bayes classifier}.
\newblock In \emph{{IJCAI 2001 workshop on empirical methods in artificial
  intelligence}}, volume~3, pages 41--46, 2001.

\bibitem[Piramuthu et~al.(1994)Piramuthu, Shaw, and
  Gentry]{piramuthu1994classification}
Selwyn Piramuthu, Michael~J Shaw, and James~A Gentry.
\newblock {A classification approach using multi-layered neural networks}.
\newblock \emph{Decision Support Systems}, 11\penalty0 (5):\penalty0 509--525,
  1994.

\bibitem[Bose et~al.(2015)Bose, Pal, SahaRay, and Nayak]{bose2015generalized}
Smarajit Bose, Amita Pal, Rita SahaRay, and Jitadeepa Nayak.
\newblock {Generalized quadratic discriminant analysis}.
\newblock \emph{Pattern Recognition}, 48\penalty0 (8):\penalty0 2676--2684,
  2015.

\bibitem[Krizhevsky et~al.(2017)Krizhevsky, Sutskever, and
  Hinton]{krizhevsky2017imagenet}
Alex Krizhevsky, Ilya Sutskever, and Geoffrey~E Hinton.
\newblock {Imagenet classification with deep convolutional neural networks}.
\newblock \emph{Communications of the ACM}, 60\penalty0 (6):\penalty0 84--90,
  2017.

\bibitem[{Huang et al.}(2017)]{8099726}
{Huang et al.}
\newblock {Densely Connected Convolutional Networks}.
\newblock In \emph{{2017 IEEE Conference on Computer Vision and Pattern
  Recognition (CVPR)}}, pages 2261--2269, 2017.
\newblock \doi{10.1109/CVPR.2017.243}.

\bibitem[Cohen and Welling(2016)]{cohen2016group}
Taco Cohen and Max Welling.
\newblock {Group equivariant convolutional networks}.
\newblock In \emph{{International conference on machine learning}}, pages
  2990--2999. PMLR, 2016.

\bibitem[{Alger et al.}(2018)]{10.1093/mnras/sty1308}
{Alger et al.}
\newblock {Radio Galaxy Zoo: machine learning for radio source host galaxy
  cross-identification}.
\newblock \emph{Monthly Notices of the Royal Astronomical Society},
  478\penalty0 (4):\penalty0 5547--5563, 05 2018.
\newblock ISSN 0035-8711.
\newblock \doi{10.1093/mnras/sty1308}.

\bibitem[{Szegedy et al.}(2017)]{szegedy2017inception}
{Szegedy et al.}
\newblock {Inception-v4, inception-resnet and the impact of residual
  connections on learning}.
\newblock In \emph{{Thirty-first AAAI conference on artificial intelligence}},
  2017.

\bibitem[{Sohn et al.}(2020)]{sohn2020fixmatch}
{Sohn et al.}
\newblock {Fixmatch: Simplifying semi-supervised learning with consistency and
  confidence}.
\newblock \emph{Advances in neural information processing systems},
  33:\penalty0 596--608, 2020.

\bibitem[Koch et~al.(2015)Koch, Zemel, Salakhutdinov, et~al.]{koch2015siamese}
Gregory Koch, Richard Zemel, Ruslan Salakhutdinov, et~al.
\newblock {Siamese neural networks for one-shot image recognition}.
\newblock In \emph{{ICML deep learning workshop}}, volume~2, page~0. Lille,
  2015.

\bibitem[Giles and Walkowicz(2018)]{10.1093/mnras/sty3461}
Daniel Giles and Lucianne Walkowicz.
\newblock {Systematic serendipity: a test of unsupervised machine learning as a
  method for anomaly detection}.
\newblock \emph{Monthly Notices of the Royal Astronomical Society},
  484\penalty0 (1):\penalty0 834--849, 12 2018.
\newblock ISSN 0035-8711.
\newblock \doi{10.1093/mnras/sty3461}.

\bibitem[Lochner and Bassett(2021)]{Lochner_2021}
M.~Lochner and B.A. Bassett.
\newblock {Astronomaly: Personalised active anomaly detection in astronomical
  data}.
\newblock \emph{Astronomy and Computing}, 36:\penalty0 100481, jul 2021.
\newblock \doi{10.1016/j.ascom.2021.100481}.

\bibitem[{He et al.}(2017)]{he2017mask}
{He et al.}
\newblock Mask r-cnn.
\newblock In \emph{Proceedings of the IEEE International Conference on Computer
  Vision (ICCV)}, Oct 2017.

\bibitem[Ronneberger et~al.(2015)Ronneberger, Fischer, and
  Brox]{10.1007/978-3-319-24574-4_28}
Olaf Ronneberger, Philipp Fischer, and Thomas Brox.
\newblock {U-Net: Convolutional Networks for Biomedical Image Segmentation}.
\newblock In Nassir Navab, Joachim Hornegger, William~M. Wells, and
  Alejandro~F. Frangi, editors, \emph{Medical Image Computing and
  Computer-Assisted Intervention -- MICCAI 2015}, pages 234--241, Cham, 2015.
  Springer International Publishing.
\newblock ISBN 978-3-319-24574-4.

\bibitem[{Luo et al.}(2020)]{luo2020survey}
{Luo et al.}
\newblock {A survey on deep hashing methods}.
\newblock \emph{ACM Transactions on Knowledge Discovery from Data (TKDD)},
  2020.

\end{thebibliography}

\end{document}